\documentclass{nature}
\DeclareUnicodeCharacter{2212}{\ensuremath{-}}
\usepackage{amsmath}
\usepackage{mathtools}
\usepackage[T1]{fontenc}
\usepackage[utf8]{inputenc}
\usepackage{graphicx}
\usepackage{amssymb}
\usepackage{dcolumn}
\usepackage{color}
\usepackage{bm}
\usepackage[normalem]{ulem}
\usepackage{bibunits}
\usepackage{hyperref}
\usepackage[capitalise]{cleveref}
\usepackage[nice]{nicefrac}
\usepackage{url}
\usepackage[labelfont=bf]{caption}
\usepackage{verbatim}
\usepackage{soul}
\usepackage{float}


\begin{document}
\begin{bibunit}

\title{Van Hove singularity-driven giant Nernst signal in twisted double bilayer graphene}

\author{Ujjal Roy$^{1}$, Monosij Roy$^{1}$, Unmesh Ghorai$^{2}$, Arkaprava Mukherjee$^{3}$, Ravi Kumar$^{4}$, Kenji Watanabe$^{5}$, Takashi Taniguchi$^{5}$, Nandini Trivedi$^{3}$, Rajdeep Sensarma$^{6}$, Subroto Mukerjee$^{1}$ and Anindya Das$^{1}$\footnote{anindya@iisc.ac.in}}
\date{}

\maketitle

\begin{affiliations}
\item Department of Physics, Indian Institute of Science, Bangalore, 560012, India.
\item School of Physics and Astronomy, Tel-Aviv University, Tel Aviv, 6997801, Israel.
\item Department of Physics, The Ohio State University, Columbus, OH 43210, USA
\item Department of Condensed Matter Physics, Weizmann Institute of Science, Rehovot 76100, Israel.
\item National Institute of Material Science, 1-1 Namiki, Tsukuba 305-0044, Japan.
\item Department of Theoretical Physics, Tata Institute of Fundamental Research, Mumbai 400005, India

\end{affiliations}

\begin{abstract}
\noindent\textbf{Twisted graphene layers host van Hove singularities (vHSs), peaks in the electronic density of states, thought to drive exotic phases in moir\'e materials, but their effect on thermal transport has remained unclear. Here we show that vHSs in twisted double bilayer graphene (tDBLG) generate an unusually large Nernst signal-the transverse voltage produced by a longitudinal temperature gradient in a magnetic field. The pronounced Nernst peaks at the vHSs of the conduction and valence bands of tDBLG are tunable by an electric field with a maximum value of $\sim 40$ $\mu V K^{-1} T^{-1}$ at $\sim 1$ $K$, which is comparable to the best-known Nernst materials. Our theoretical calculations show that the large enhancement of the Nernst signal arises from the Lifshitz transitions around the vHSs. These findings establish the Nernst effect as a sensitive probe of Fermi-surface topology in moir\'e materials, and identify a universal thermoelectric signature of van Hove singularities.}
\end{abstract}

\noindent\textbf{Introduction.} In recent years, twisted double bilayer graphene (tDBLG) has emerged as a versatile platform for exploring correlated and topological phases in moir\'e materials \cite{Andrei2021}. Unlike twisted bilayer graphene \cite{Andrei2020-ad}, where correlated phenomena are restricted to a magic angle ($\simeq 1.1^\circ$), tDBLG provides an additional in situ tuning knob, out-of-plane displacement field ($D$) \cite{Jung2019,Koshino2019} to control various emergent phases such as correlated insulators \cite{Cao2020,PKim2020,Shen2020}, spin, valley-polarized states \cite{yankowitz2021,Shen2020}, and superconductivity \cite{Su2023,Ashvin2019}. The single particle electronic structure of tDBLG is characterized by prominent van Hove singularities (vHS) arising from saddle points in the moir\'e minibands, which are tunable with $D$. These singularities coincides with Lifshitz transitions and are widely believed as precursors to electronic instabilities like cascaded transition \cite{Yazdani2020,Zondiner2020}, superconductivity \cite{Kohn1965,Gonzalez2008-rw,Nandkishore2012-cv} and itinerant magnetism \cite{Fleck1997-yo,Betouras2018,He2019}. Despite their central role, experimental access to vHS has remained largely confined to electrical transport \cite{Andrei2010,Pasupathy2019,Nadj2019,Wu2021,Mak2025}. Its consequences for thermal transport and in particular for the transverse thermoelectric response or `Nernst' effect remain unexplored. Since the Nernst effect is sensitive to Fermi surface topology, Berry curvature, and superconducting fluctuations, making it a powerful probe of studying correlated moir\'e systems.

\noindent The Nernst effect \cite{Behnia2009,Behnia2016-es} $(N = - S_{yx}= -E_{y}/\nabla_{x} T)$ is the thermoelectric analogue of the Hall effect, referring to the generation of a transverse electric field (Fig. 1a) by a longitudinal temperature gradient in the presence of a perpendicular magnetic field. 
For normal quasiparticles, as depicted by Fig. 1b, the hot and cold carriers above and below the chemical potential, $E_F$, diffuse in opposite directions under a thermal gradient; and in a magnetic field, these counter-propagating carriers are deflected in opposite transverse directions (Fig. 1d), and cancel each other. Using semiclassical Boltzmann transport, one can show that this cancellation is exact (which is known as Sondheimer cancellation \cite{Sondheimer1948-ff}) only if the carrier mobility ($\mu$) is energy ($\epsilon$) independent. Any finite Nernst signal and its sign will depend on $\sim \partial\mu/\partial\epsilon$ around $E_F$. Since in metal $\partial\mu/\partial\epsilon$ is negligibly small, and thus Nernst signal is scarce. Historically, the Nernst effect gained attention with the discovery of a large signal in bismuth \cite{Behnia2007} at low temperatures, still among the largest reported, and attributed to its high mobility and small Fermi energy (large $\partial\mu/\partial\epsilon$). The effect later became central to high-temperature superconductivity \cite{Ong2000-os,ong2006,Behnia2016-es}, where large signal originates from vortex dynamics and superconducting fluctuations \cite{Pourret2006-pq,Huse2002}. 
Further, 
Nernst signal have also been observed in semimetals \cite{Behnia2010,Behnia2015wte2} (e.g., $WTe_2$, graphite) and narrow-gap semiconductors \cite{Ong2013,Behnia2015,BiTei2015}, arising from multiband (ambipolar) transport \cite{Behnia2003}. A photo-Nernst effect near a vHS has been reported under optical excitation in hBN-graphene moir\'e superlattices\cite{sanfengWu2016}. However, there are no equilibrium transport studies on the Nernst signal at vHS of any twisted platform.


\begin{figure*}[htbp]
\centerline{\includegraphics[width=1\textwidth]{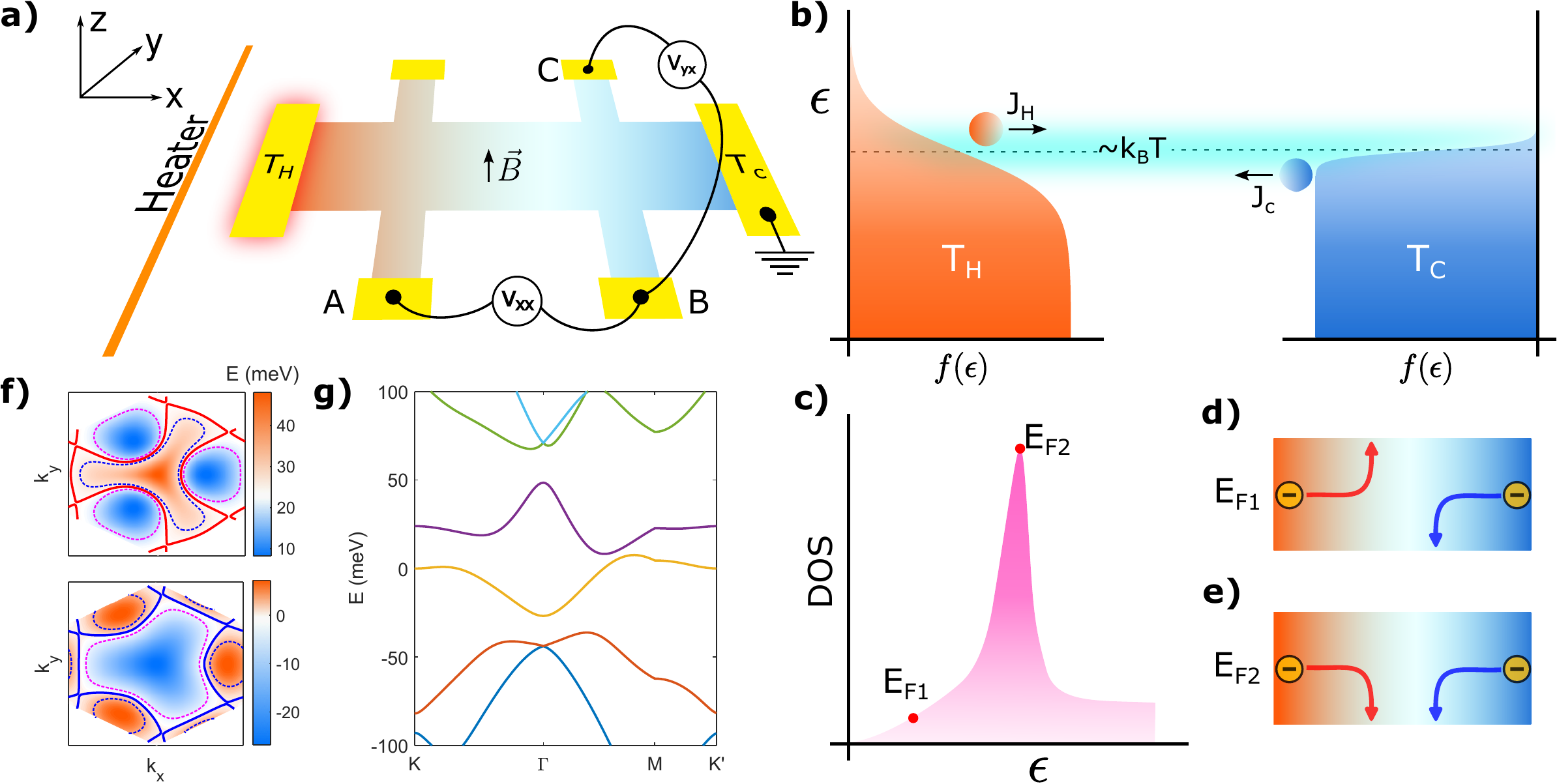}}
\caption{\textbf{Measurement schematic, Lifshitz
transitions and thermally driven carrier motion in magnetic field. (a)} Schematic of the thermoelectric measurements. Passing current through the metallic heater creates a longitudinal temperature gradient, and the resulting open-circuit thermoelectric voltages are measured between contacts A-B ($V_{xx}$) and B-C ($V_{yx}$), corresponding to the $S_{xx}$ ($-V_{xx}/\Delta T$) and $S_{yx}$ ($-\frac{V_{yx}}{\Delta T}\frac{L}{W}$), respectively (see SI Fig. 17 for details). The vertical arrow indicates the direction of the applied perpendicular magnetic field ($B$). \textbf{(b)} Thermally driven carrier diffusion between the hot and cold end. At steady state, high-energy carriers diffuse toward the cold end, while low-energy carriers move toward the hot end. The corresponding current contributions, labeled as $J_H$ and $J_C$, and satisfy $J_H+J_C=0$ (open circuit condition). \textbf{(c)} A schematic representation of the DOS as a function of energy, where the peak mimics the presence of a vHS. \textbf{(d)} Illustration of carrier motion in the presence of $\Vec{B}$. The thermally driven carrier motion described in (b) is deflected transversely by the Lorentz force. When the chemical potential lies away from the vHS (denoted $E_{F1}$ in (c)), the deflections of high- and low-energy carriers are in opposite transverse directions. \textbf{(e)} Carrier motion when the chemical potential is tuned to the vHS ($E_{F2}$ in (c)). Since the effective mass changes sign across the saddle point, causes both the high- and low-energy carriers to deflect toward the same side. \textbf{(f)} Top panel, Energy contours around the Lifshitz transition of $1.66^0$ tDBLG for the conduction band at zero perpendicular electric field. Three Fermi pockets merge to a single pocket. Same happens for the valency band (bottom panel). Colorbar represents the Fermi energy. \textbf{(g)} Band dispersion along the high symmetry direction.}
\label{fig1}
\end{figure*}

\noindent Here, we have carried out the first Nernst measurements on 
tDBLG with twist angles of $1.66^0$ (D1) and $1.20^0$ (D2). From the Hall response, we map the position of the vHS in both the conduction and valence bands as a function of $D$, and it agrees well with our theoretically calculated density of states (DOS) based on the continuum model\cite{Koshino2019,Mohan2021,Jung2019}. In the presence of a longitudinal temperature gradient, the longitudinal thermopower, $S_{xx}$ ($E_{x}/\nabla_{x} T$), changes sign across the vHS, and vanishes at the vHS due to particle-hole symmetry. In contrast, the Nernst response, $S_{yx}$, exhibits large positive peaks around the vHS for both the conduction and valence bands, and is tunable with $D$ with a maximum magnitude of $\simeq 20$ $\mu V K^{-1} T^{-1}$ and $\simeq 40$ $\mu V K^{-1} T^{-1}$ at $\simeq 1$ $K$ for $1.66^0$ and $1.20^0$, respectively. These values remain comparable to heavy fermion metal ($PrFe_4P_{12}$)\cite{Behnia2009} at such a lower temperature, and only bismuth has higher values than these. Apart from vHS, the $S_{yx}$ remains negligibly small throughout the bandwidth of the tDBLG. Based on semiclassical Boltzmann transport equations, we calculate $S_{yx}$ using the measured electrical conductivity tensor and theoretical DOS, which matches well with our measured $S_{yx}$. 
We show that both the hot and cold carriers in tDBLG are deflected towards the same direction, as illustrated in Fig. 1e, in contrast to opposite directions in metal (Fig. 1d), due to Lifshitz transition around the vHS, and hence produce a large Nernst signal. The sign of $S_{yx}$ remains positive 
for both the vHSs in the conduction and valence bands, and we argue that this positive sign of the $S_{yx}$ is universal around a vHS in contrast to metals or semimetals.

\begin{figure*}[htbp]
\centerline{\includegraphics[width=1\textwidth]{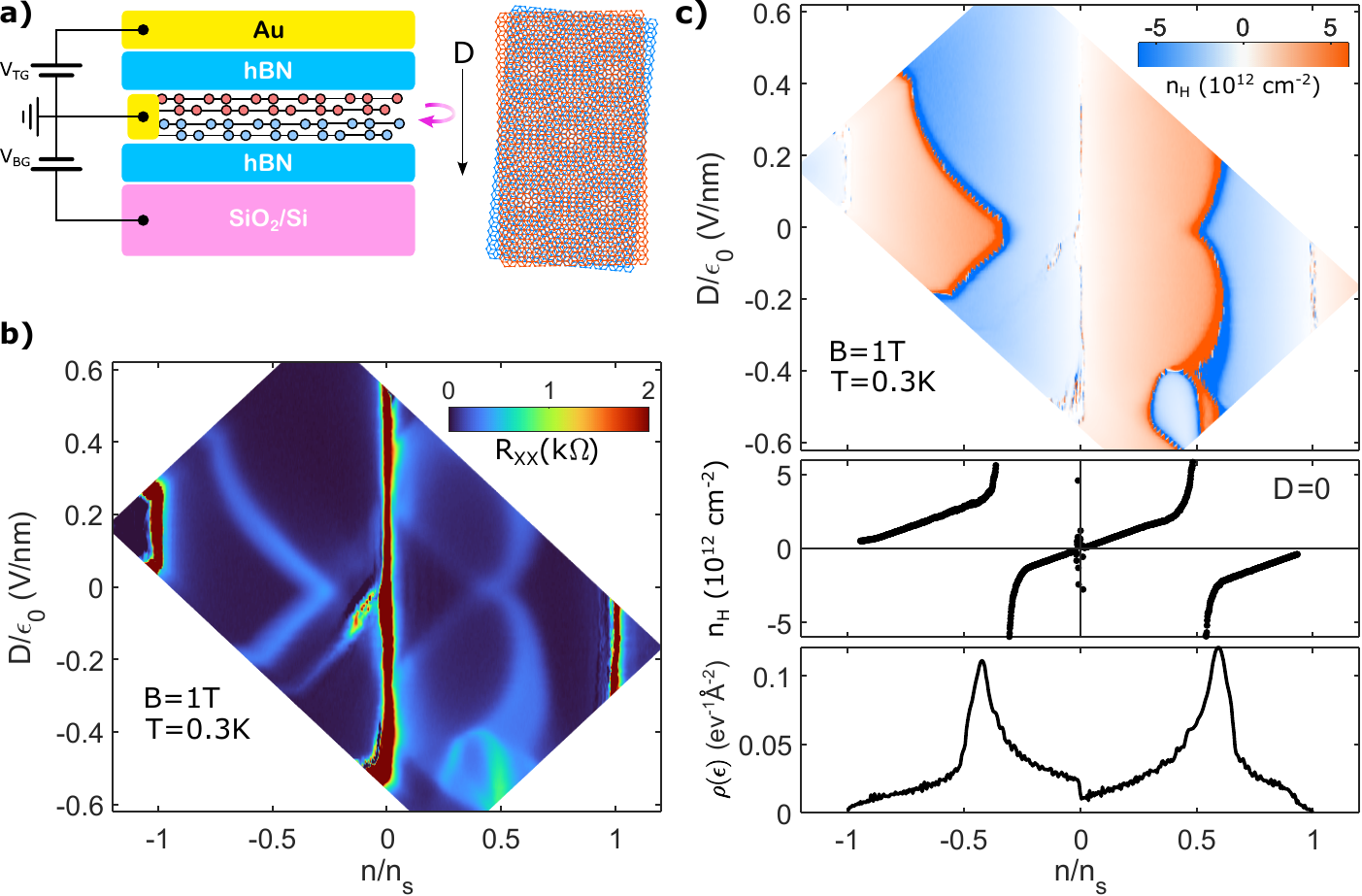}}
\caption{\textbf{Resistance and Hall response tracing vHS. (a)} (Left) Schematic cross-section of the hBN-encapsulated tDBLG stack. (Right) The moir\'e heterostructure is formed by stacking two Bernal BLG flakes with a 
twist angle. The TG and the BG are biased by external voltage sources ($V_{TG}$ and $V_{BG}$), allowing for independent control of $n$ and $D$ for device D1. The arrow indicates the positive direction of $D$. \textbf{(b)} 2D colormap of 4-probe longitudinal resistance, $R_{xx}$, measured at $B=1T$ and $T=0.3K$, as a function of normalized carrier density $n/n_s$ and $D$. \textbf{(c)} (Top) Evolution of Hall density, $n_H=B/eR_{xy}$, (here $e$ is the electronic charge and $R_{xy}$ is the Hall resistance) in the same $n-D$ plane. (Middle) Line cut of $n_H$ for $D=0$. (Bottom) Calculated single particle DOS at zero $D$ as a function of $n/n_s$. The prominent peaks denote the position of the vHS, which align with the $n_H$ divergences in the middle panel, identifying the locations of the Lifshitz transitions.}
\label{fig2}
\end{figure*}

\noindent\textbf{Results.}
\newline\noindent\textbf{Device and working principle.} 
\noindent Our tDBLG samples are hBN encapsulated and prepared using the standard cut and stack transfer method. The $SiO_2/Si$ substrate serves as the backgate (BG), while the topgate (TG) consists of a metal layer deposited on the top dielectric, as illustrated in \cref{fig2}a. This allows independent control of carrier density (n) and displacement field (D) (see the Methods section for details). The resulting stack is patterned into a Hall bar configuration. To facilitate thermoelectric measurements, a metallic heater is positioned to generate a longitudinal temperature gradient along the Hall bar direction, as depicted in \cref{fig1}a. We utilize the well-established $2\omega$ technique for these measurements\cite{kim2009,Paul2022-sn}. In the presence of a perpendicular magnetic field ($B$), the resulting thermoelectric voltages (longitudinal and transverse) are measured as shown in \cref{fig1}a (see section 10 of the supplementary information (SI) for more details). The temperature gradient is determined through a calibration procedure based on the resistance variation of the tDBLG device measured both as a function of heater excitation and the bath temperature. Further details of the calibration method are provided in SI section 11.

\noindent\textbf{Electrical response.} \cref{fig2} presents the longitudinal resistance ($R_{xx}$) and Hall resistance ($R_{xy}$) at $B=1T$ and $T=300mK$ using standard lock-in techniques. 
\cref{fig2}b shows the 2D colormap of the symmetrized $R_{xx}$ as a function of the normalized carrier density, $n/n_s$ and $D$. Here, $n_s\approx6.5\times10^{12}$ $cm^{-2}$ corresponds to the carrier density required to fully fill the moir\'e bands, yielding a twist angle of $\theta\approx1.66^\circ$ (see SI section 2 for the twist angle determination). In addition to the highly resistive states at charge neutrality and full filling, weaker resistive features emerge 
in both the valence $(n<0)$ and conduction $(n>0)$ bands. The corresponding carrier densities 
shift systematically with 
$D$. The 2D colormap of the Hall density ($n_H$) extracted from the measured anti-symmetrized $R_{xy}$  is shown in \cref{fig2}c-top panel, where the divergence of Hall density, followed by a sign change along the $n-D$ trajectory, indicate the position of the vHS arising from the Lifshitz transition. Such behavior is consistent with earlier reports and is well captured by our theoretically calculated band dispersion using continuum model\cite{Koshino2019,Mohan2021,Jung2019}. The energy contour plots for the conduction (top panel) and valence band (bottom panel) for $D=0$ are shown in \cref{fig1}f. For the conduction band, as energy is increased, at a critical energy, the isolated neighboring three 
Fermi pockets (blue-shaded region) expand, touch and eventually merge, marking a transition to a single 
Fermi pocket around the $\Gamma$ point (red shaded region). Similar thing happen for the valence band. 
The critical energies, where the Lifshitz transitions happens, are the positions of the vHSs. In Fig. 1g, the band dispersion along the high-symmetry directions is shown for $D=0$, and the extracted DOS is plotted in the Fig. 2c-bottom panel, where the large DOS peaks appear at the Lifshitz transitions, and matches well with the position of density at which $n_H$ diverges and changes sign as shown in Fig. 2c-middle panel (a cutline of $n_H$ at $D =0$). Note that in Fig. 2b and 2c, at higher negative $D \sim -0.4V/nm$ to $-0.6V/nm$, a weak signature of correlation near the half-filling emerges.

\begin{figure*}[htbp]
\centerline{\includegraphics[width=1\textwidth]{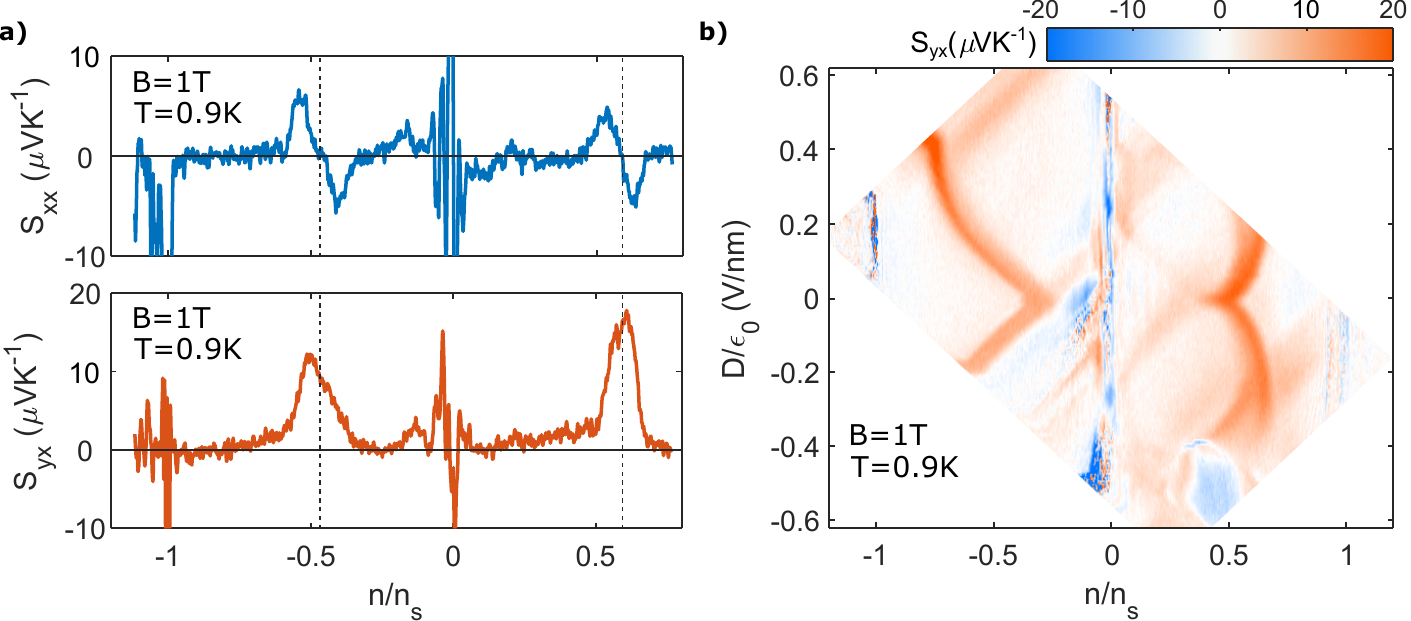}}
\caption{\textbf{Thermopower and Nernst response. (a)} Line cuts of the thermoelectric coefficients (TEC), $S_{xx}$ (blue) and $S_{yx}$ (orange) as a function of $n/n_s$ for a fixed displacement field, $D = 0.1 Vnm^{-1}$, and $B = 1T$ and $T=0.9K$. $S_{xx}$ changes sign across the vHS and becomes zero at vHS, whereas the $S_{yx}$ exhibits positive peak around the vHS. \textbf{(b)} 2D colormap of the Nernst signal, $S_{yx}$ as function of $n/n_s$ and $D$, for $B=1T$ and $T=0.9K$. Enhanced $S_{yx}$ values are observed along the same $n$-$D$ trajectories as the vHS identified in Hall measurement, \cref{fig2}b. and c.} 
\label{fig3}
\end{figure*}

\noindent\textbf{Thermopower and Nernst response.} 
\cref{fig3}a (top-panel) shows longitudinal thermopower, $S_{xx}$ as a function of $n$ at $D=0.1V/nm$ for $T = 0.9K$ and $B = 1T$. The 2D colormap of $S_{xx}$ with $n-D$ shown in SI Fig. 6a. As expected,\cite{Mott2012-kk,Paul2022-sn}, the $S_{xx}$ changes sign around the charge neutrality point, CNP ($n/n
_S =0$), and moire band full-filling ($n/n
_S =\pm 1$), which can be clearly seen at a slightly higher temperature $\sim3K$ as shown in SI Fig. 7a. However, at lower temperatures, since the magnitude of the signal decreases, the effect of disorder dominates at the low density of states like CNP and moire gap. In contrast, even at lower temperature (as seen in Fig. 3a), the $S_{xx}$ exhibits a robust positive peak to the left of the vHS and a negative peak to the right, crossing through zero at the vHS. These features precisely track the trajectory of the Lifshitz transitions as identified in the Hall data in Fig. 2c. The sign reversal across the vHS re-establishes the change in the Fermi surface topology. 

\noindent \cref{fig3}a-bottom panel, presents the the Nernst response, $S_{yx}$ as a function of $n$ at $D=0.1V/nm$ for $T = 0.9K$ and $B = 1T$. The 2D colormap of $S_{yx}$ with $n-D$ shown in \cref{fig3}b. Other than irregular kinks at the CNP and moire gap, $S_{yx}$ exhibits pronounced positive peak around the vHS, whereas it remains almost negligibly small throughout the bandwidth of the tDBLG. Comparing $S_{xx}$ and $S_{yx}$ in Fig. 3a, it becomes evident that while $S_{xx}$ changes its polarity across the vHS, $S_{yx}$ manifests a large positive peak at vHS, which traces the position of the vHS as a function on $n-D$, and remain identical to $n_H$ plot (Fig. 2c top panel). Note that all thermoelectric voltages have been symmetrized (for longitudinal components) and anti-symmetrized (for transverse components) to eliminate any cross-mixing between the signals (see SI Fig. 5). The linearity of the Nernst signal with $B$ was verified and shown in SI Fig. 9 (for device D1) and 11 (device D2). 

\noindent Why the $S_{yx}$ gives rise to large postive peak at vHS? First, we will present an intuitive picture, whereas in the next section, we will compare with the calculated $S_{yx}$ using semiclassical Boltzmann transport. A temperature gradient induces an asymmetry in the Fermi distribution, driving carriers to flow. At steady state, there is no electrical current flows since thermopower measurement is performed in open-circuit condition. However, there will be heat current flowing from the hot end to the cold end, and can be visualized as two counter-propagating electrical currents flowing around the chemical potential as shown in Fig. 1b; current from the high-energy carriers from hot to cold end, and vice-versa from the low-energy carriers. In presence of a perpendicular $B$, these thermally driven carriers will be deflected transversely by the Lorentz force. When the chemical potential is away from a vHS, the deflections of high- and low-energy carriers are in opposite transverse directions as shown in Fig. 1d. The resultant $S_{yx}$ depends on their cancellation, and the cancellation is perfect when the mobilities of the high- and low-energy carriers are equal, resulting in a zero Nernst signal\cite{Sondheimer1948-ff}. However, at the vHS, since the the effective mass changes sign across the saddle point, the high-energy carriers will be deflected opposite to the previous case (away from vHS), causing both high- and low-energy carriers to deflect toward the same downward side as shown in Fig. 1e. This results in enhanced carrier accumulation and a large positive $S_{yx}$. 

\noindent As shown in SI-Fig. 17 and SI-Fig. 18, for the valence band also, the $S_{yx}$ gives rise to positive signal (SI-Fig. 18b). This behavior is unique to the vHSs, unlike in conventional metals, where the Nernst signal is governed by the energy derivative of the carrier mobility at the Fermi level, and depending on that the sign can be positive or negative. 
Note that, we observe a negative $S_{yx}$ at the CNP and moir\'e gap, clearly visible from $T=3K$ onward (SI-Fig. 10), and are also measured in bilayer graphene samples (SI-Fig. 15) around the Dirac point. These observations remain consistent with the semiclassical description as shown in SI-Fig. 18. 
However, around the Dirac point, we observe negative $S_{yx}$ only at higher temperatures, whereas large positive $S_{yx}$ was measured at vHSs at lower temperatures. This is the importance of vHSs, even though the $k_BT$ window is smaller at lower temperature, but due to the larger magnitude of $J_H$ and $J_C$ (Fig. 1d) coming from the enhanced DOS, produce large $S_{yx}$. 

\begin{figure*}[htbp]
\centerline{\includegraphics[width=1\textwidth]{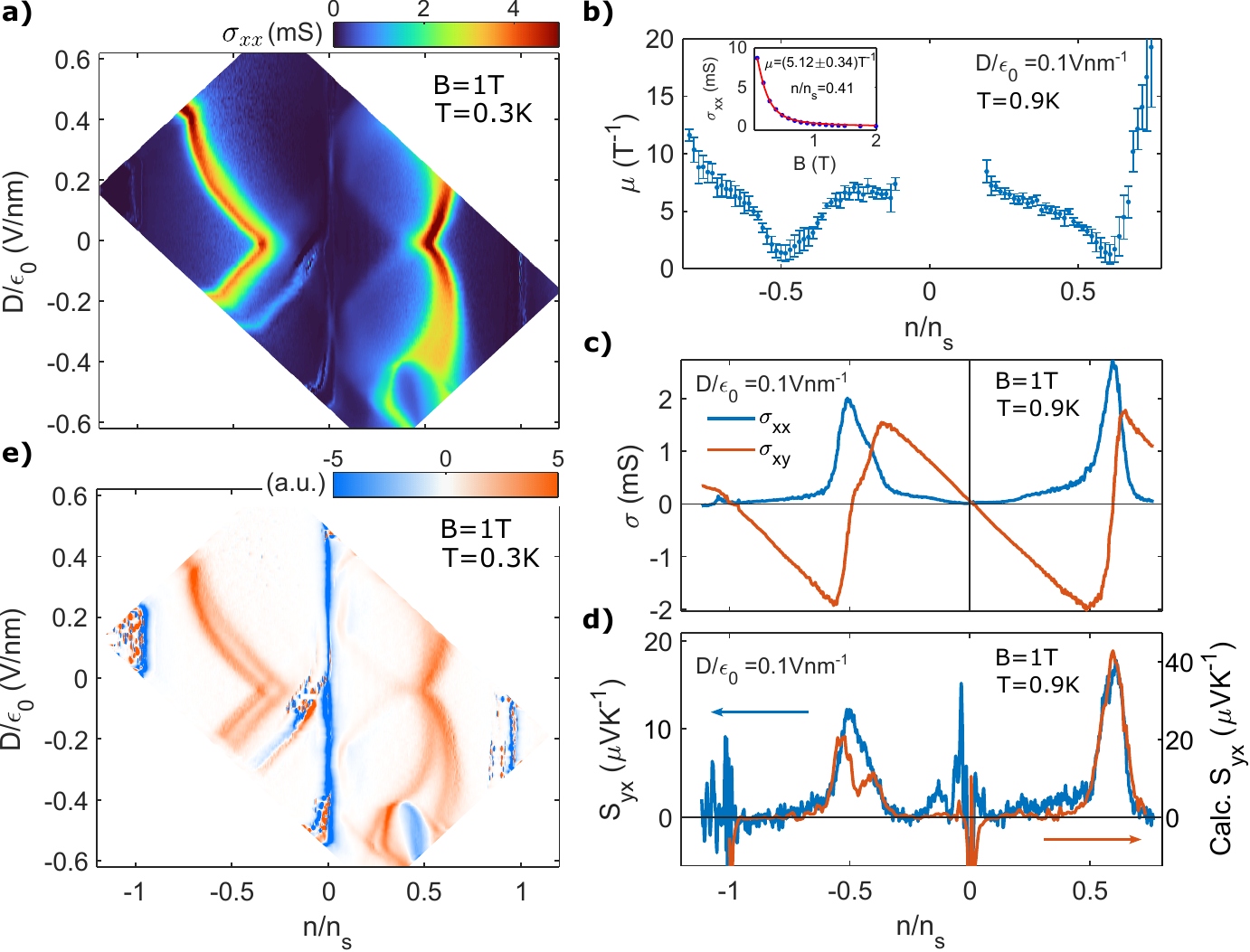}}
\caption{\textbf{Conductivity, mobility and Nernst from semiclassical theory. (a)} 2D colormap of $\sigma_{xx}$ with $n - D$ at $B = 1T$ and $T = 0.3K$. \textbf{(b)} Extracted Hall mobility as a function of carrier concentration at $D/\epsilon_0=0.1Vnm^{-1}$. Error bar represent the fitting error during the extraction of $\mu$. Inset shows the magnetic field dependence of $\sigma_{xx}$ (solid circles) at $n/n_s=0.41$. The Red solid line is the fitting to extract the mobility. \textbf{(c)} $\sigma_{xx}$ and $\sigma_{xy}$ cut-lines with $n/n_s$ at $D/\epsilon_0=0.1Vnm^{-1}$. The $\sigma_{xx}$ shows peak around the vHS whereas the $\sigma_{xy}$ changes sign. \textbf{(d)} Using the above conductivity and theoretical DOS, the $S_{yx}$ is calculated using \cref{eq1}, and shown in orange solid line. In comparison, the measured $S_{yx}$ is shown in blue solid line. \textbf{(e)} 2D colormap of calculated $S_{yx}$ using \cref{eq1}, without the DOS contribution.}
\label{fig4}
\end{figure*}

\noindent\textbf{Nernst from semiclassical Boltzmann transport.} In linear response, the thermoelectric voltage generated due to a longitudinal temperature gradient is $\mathbf{E_{th}}=\bar\sigma^{-1}.\bar\alpha.\nabla_x T$, where $\bar\sigma$ and $\bar\alpha$ are the conductivity and thermoelectric tensors in presence of a magnetic field, and will have longitudinal, $S_{xx} = E_x/\nabla_x T = \frac{\alpha_{xx}\sigma_{xx}+\alpha_{xy}\sigma_{xy}}{\sigma_{xx}^2+\sigma_{xy}^2}$ and transverse, $S_{yx} = E_y/\nabla_x T = \frac{\alpha_{xx}\sigma_{xy}-\alpha_{xy}\sigma_{xx}}{\sigma_{xx}^2+\sigma_{xy}^2}$ components. Following the semiclassical description, the $\bar\alpha$ is expressed in terms of $\bar\sigma$ as, $\bar\alpha=-\frac{\pi^2k_B^2T}{3e}\frac{\partial\bar\sigma}{\partial\epsilon}$, and $S_{yx}$ can be written in terms of only the electrical conductivity terms as:
\begin{equation}
    S_{yx}=\frac{\pi^2}{3}\frac{k_B^2T}{e}\cos^2\Theta_H\frac{\partial\tan\Theta_H}{\partial \epsilon}|_{\epsilon = \epsilon_{F}} = \frac{\pi^2}{3}\frac{k_B^2T}{e}\cos^2\Theta_H\frac{\partial\tan\Theta_H}{\partial n}\rho (n)\bigg|_{\epsilon = \epsilon_{F}}
    \label{eq1}
\end{equation}
Where $\Theta_H$ is the Hall angle ($\tan\Theta_H = \frac{\sigma_{xy}}{\sigma_{xx}} = \mu B$), and $\rho(n)$ is the DOS. This was developed by Oganesyan and Ussishkin \cite{Ussishkin2004} and 
can be reduced to $S_{yx}=\frac{\pi^2}{3}\frac{k_B^2T}{e}\frac{\partial\tan\Theta_H}{\partial \epsilon}|_{\epsilon = \epsilon_{F}} = \frac{\pi^2}{3}\frac{k_B^2T B}{e}\frac{\partial \mu}{\partial \epsilon}|_{\epsilon = \epsilon_{F}}$ for a small Hall-angle approximation and typically satisfies for metal, and thus the sign and magnitude of the Nernst signal depend on how mobility varies with energy around the Fermi energy. However, for our tDBLG, we show below that $\tan\Theta_H$, ie $\mu B>1$, and thus the small Hall-angle approximation cannot be used. 

\noindent \cref{fig4}a shows $\sigma_{xx}$ with $n-D$ at $B=1T$ and $T=0.3K$, and shows a peak along the vHS trace. 
To understand, we extract $\mu$ from $B$ dependence of $\sigma_{xx}$, by fitting the expression $\sigma_{xx}=\frac{\sigma_0}{1+\mu^2B^2}$ (where $\sigma_0$ is the conductivity at $B=0$), as shown in the inset of \cref{fig4}b for $n/n_s=0.41$ and $D = 0.1V/nm$. \cref{fig4}b shows $\mu$ as a function of density (see SI Fig. 12a for $\sigma_{xx}$ with $n$ for different $B$ for mobility extraction, and can be seen that the data is well fitted except around the CNP and fulfilling). Key observations are: $\mu B \ge 1$ throughout the phase space, and $\mu$ varies significantly with $E_F$, with minimum value around the vHS, thus giving rise to peak in $\sigma_{xx}$. A cut-line of $\sigma_{xx}$ from Fig. 4a for $D=0.1$ $Vnm^{-1}$ is shown in Fig. 4c with the corresponding $\sigma_{xy}$. Utilizing $\sigma$ and theoretically calculated DOS, we calculate the $S_{yx}$ using \cref{eq1}, which is shown in Fig. 4d (orange line) in comparison with the experimentally measured $S_{yx}$ (blue line). The agreement between them is quite remarkable, though the magnitude of the measured signal is on the  smaller side, which could be due to uncertainty in DOS, since the real device with angle inhomogeneity can broaden the DOS, which is 
not taken into account 
in the theoretical calculation. A better insight of Eq. 1 can be seen in SI-Fig. 12b, where both the $\cos\Theta_H$ and $\tan\Theta_H$ are plotted as a function of carrier density. Finally, \cref{fig4}e, shows the 2D colormap of \cref{eq1} (without the DOS contribution). Notably, this calculation reproduces the prominent peak features observed in the experimental Nernst data shown in \cref{fig3}b. In SI section 6, we show the calculated $S_{xx}$ using semiclassical equation (Mott formula), which matches well with the measured $S_{xx}$.

\noindent \textbf{Discussion.} The non-interacting semiclassical description captures our data nicely, which further can be seen from the comparison (see SI Fig. 13) of thermoelectric tensors, $\alpha_{xx}$ and $\alpha_{xy}$ obtained 
from the experimentally measured $S_{xx}$, $S_{yx}$, $\sigma_{xx}$ and $\sigma_{xy}$ with the extracted $\bar\alpha$ from the conductivity tensor using $-\frac{\pi^2k_B^2T}{3e}\frac{\partial\bar\sigma}{\partial\epsilon}$. Around the vHSs, the $\alpha_{xx}$ changes sign reconfirming the changes of the Fermi surface topology, whereas $\alpha_{xy}$ exhibits large negative peaks.

\noindent So far, we have established the connection between the thermoelectric tensors and electrical conductivity using the semiclassical Mott relation. 
Further, we develop a minimal non-interacting two-band tight-binding model on a triangular lattice, 
having the vHS and the associated Lifshitz transitions (see section SI-12). 
We theoretically calculate the electrical conductivity ($\sigma_{xx}$ and $\sigma_{xy}$) and thermoelectric coefficients ($S_{xx}$ and $S_{yx}$), which exhibits 
large postive Nernst peak as a thermoelectric signature of the vHS associated with the Lifshitz transition. 

\noindent \textbf{Conclusion.} In summary, we show that van Hove singularities in tDBLG produce a giant, universally positive Nernst signal that precisely tracks Lifshitz transitions. A semiclassical Boltzmann analysis reproduces this response quantitatively, tracing it to the effective-mass sign reversal at the saddle point - a mechanism distinct from conventional metals. Since this arises purely from band topology, we expect giant vHS-driven Nernst signals to be a generic feature across moir\'e and flat-band materials, establishing the Nernst effect as a powerful new probe of Fermi-surface topology.

\noindent\textbf{Methods.} \\
\noindent\textbf{Device fabrication.} To fabricate the twisted graphene heterostructure, we employed the standard "cut-and-stack" technique following our previous work\cite{Paul2022-sn}. An exfoliated bilayer graphene (BLG) flake on a $SiO_2$ substrate was precisely bisected using a sharp tungsten tip; the remainder of the stacking procedure follows that described in our previous work. Ohmic contacts and local heater were then defined by electron beam lithography (EBL), followed by reactive ion etching (RIE) with a $CHF_3/O_2$ plasma and thermal evaporation of Cr/Pd/Au (of thickness $5/12/70$ nm). A metal top-gate was then integrated through a secondary EBL and metallization step. Finally, the complete heterostructure was patterned into a Hall-bar geometry using a final EBL step and plasma etching.\\
\noindent\textbf{Measurements.} All the measurements were carried out in a $^3He$ cryostat, equipped with a superconducting magnet, at a base temperature of $\sim250mK$. Transport measurements were performed using standard low frequency lock-in techniques. Four-terminal resistance measurements were conducted with a constant current bias of $\sim5nA$ and thermoelectric voltages were acquired using the standard $2\omega$ method (see the SI section 10 for details). In a finite magnetic field, the longitudinal transport components ($V_{xx}$) were symmetrized and the transverse components $(V_{yx})$ were anti-symmetrized with respect to field direction as:
$$V_{xx}=\frac{V_{xx}(B^+)+V_{xx}(B^-)}{2}$$
$$V_{yx}=\frac{V_{yx}(B^+)-V_{yx}(B^-)}{2}$$
where $V(B^+)$ and $V(B^-)$ denote the values measured under positive and negative magnetic fields, respectively. 
\\
\noindent The $n-D$ plots were obtained by sweeping the carrier density $n$ at fixed displacement field $D$, achieved by simultaneously varying the top-gate (TG) and bottom-gate (BG) voltages according to:
$$D=\frac{C_{TG}(V_{TG}-V_{TG0})-C_{BG}(V_{BG}-V_{BG0})}{2}$$
$$ n=C_{TG}(V_{TG}-V_{TG0})+C_{BG}(V_{BG}-V_{BG0})$$
\noindent Here, $V_{BG}$ and $V_{TG}$ are the applied top and bottom gate voltage, respectively; $C_{BG}$ and $C_{TG}$ are the respective gate capacitances per unit area, and $V_{BG0}$ and $V_{TG0}$ denote the voltage offsets required to reach the charge neutrality point (CNP) for each gate.

\subsection{Acknowledgements}
A.D. thanks the Anusandhan National Research Foundation (ANRF) with project no: SP/ANRF-26-0244 and Department of Science and Technology (DST/NM/TUE/QM-5/2023) for the financial support. K.W. and T.T. acknowledge support from the Elemental Strategy Initiative conducted by the MEXT, Japan and the CREST (JPMJCR15F3), JST. 

\subsection{Author contributions}
U.R. contributed to device fabrication, measurement, data acquisition and analysis. M.R. and R.K. contributed to the measurement and data analysis. U.G. and R.S. contributed to the theoretical framework of the continuum model for tDBLG. A.M. and N.T. contributed to the minimal model-based theoretical calculations. S.M. contributed to understanding the Nernst effect. K.W. and T.T. synthesized the hBN single crystals. All the authors contributed to writing the manuscript.

\subsection{Competing financial interests}
The authors declare no competing financial interests.

\subsection{Data availability}
The data presented in the manuscript are available from the corresponding author upon request.

\noindent\textbf{References}
\putbib[ref]
\end{bibunit}
\clearpage

\begin{bibunit}[naturemag]
\setcounter{page}{1}
\setcounter{section}{0}
\setcounter{figure}{0}
\setcounter{table}{0}
\setcounter{equation}{0}

\renewcommand{\thefigure}{S\arabic{figure}}
\renewcommand{\thetable}{S\arabic{table}}
\renewcommand{\theequation}{S\arabic{equation}}
\renewcommand{\thesection}{S\arabic{section}}


\title{Supplementary Information: Van Hove singularity-driven giant Nernst signal in twisted double bilayer graphene}

\author{Ujjal Roy$^{1}$, Monosij Roy$^{1}$, Unmesh Ghorai$^{2}$, Arkaprava Mukherjee$^{3}$, Ravi Kumar$^{4}$, Kenji Watanabe$^{5}$, Takashi Taniguchi$^{5}$, Nandini Trivedi$^{3}$, Rajdeep Sensarma$^{6}$, Subroto Mukerjee$^{1}$ and Anindya Das$^{1}$\footnote{anindya@iisc.ac.in}}

\maketitle

\begin{affiliations}
\item Department of Physics, Indian Institute of Science, Bangalore, 560012, India.
\item School of Physics and Astronomy, Tel-Aviv University, Tel Aviv, 6997801, Israel.
\item Department of Physics, The Ohio State University, Columbus, OH 43210, USA
\item Department of Condensed Matter Physics, Weizmann Institute of Science, Rehovot 76100, Israel.
\item National Institute of Material Science, 1-1 Namiki, Tsukuba 305-0044, Japan.
\item Department of Theoretical Physics, Tata Institute of Fundamental Research, Mumbai 400005, India

\end{affiliations}

\noindent\textbf{\Large Contents}

\begingroup
\setlength{\parskip}{1pt} 
\tableofcontents
\endgroup

\section{Optical image of the devices}

\begin{figure}[H]
\centerline{\includegraphics[width=1\textwidth]{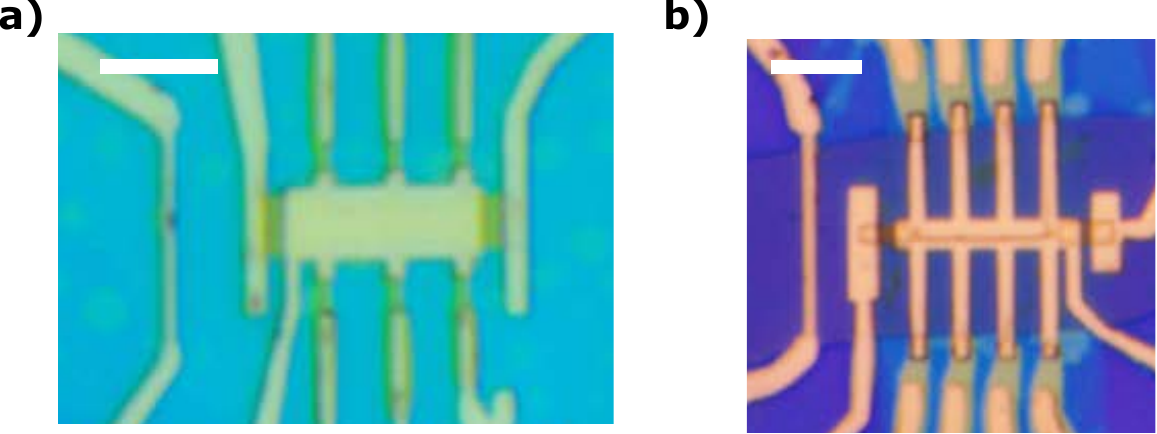}}
\caption{Optical micrograph of device D1 of twist angle $\theta_t\approx1.66^\circ$ \textbf{(a)} and D2 of $\theta_t\approx1.20^\circ$ \textbf{(b)}. D1 uses a Si back-gate, while D2 employs a graphite back-gate. Scale bar: $5\mu m$.}
\label{image}
 
\end{figure}

\section{Twist angle determination}

\begin{figure}[H]
\centerline{\includegraphics[width=1\textwidth]{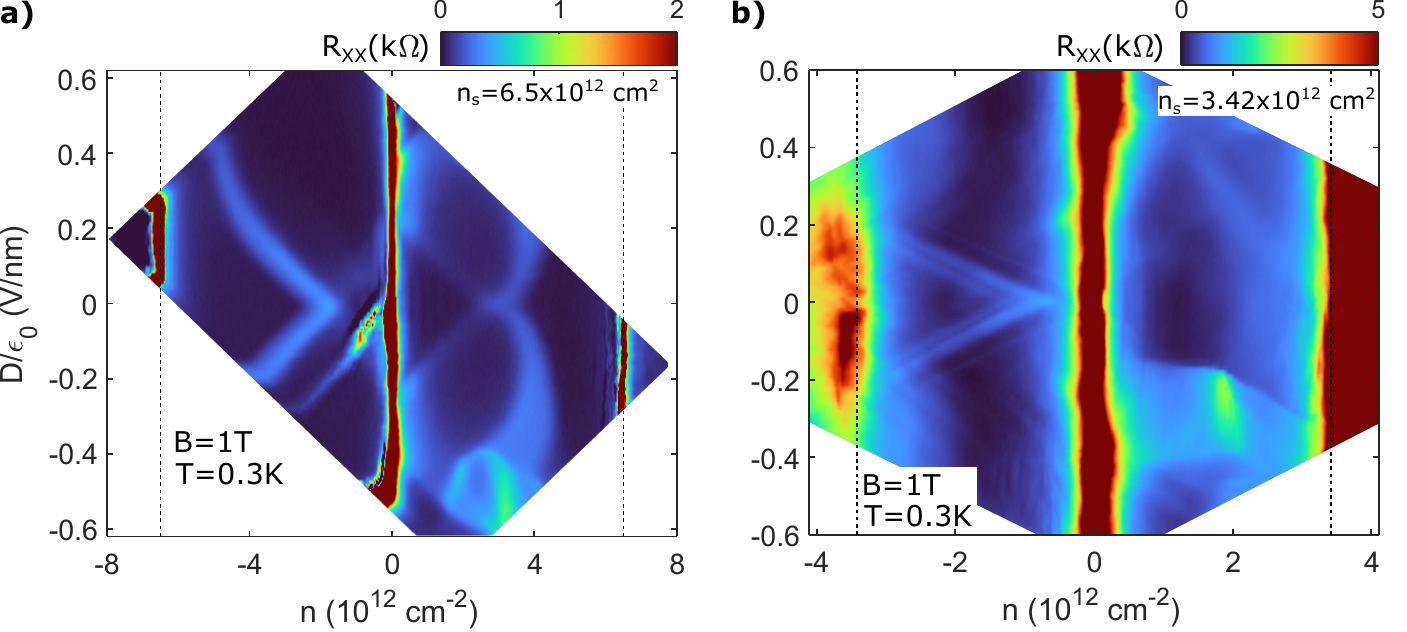}}
\caption{\textbf{Twist angle determination.} 2D colormap of $R_{xx}$ in $n-D$ plane, for device D1 with twist angle $\theta\approx1.66^\circ$ \textbf{(a)} and D2 with $\theta\approx1.20^\circ$ \textbf{(b)}. The data shows single particle gaps, indicated by the insulating resistive states at the CNP as well as superlattice densities, defined as $n=\pm n_s$, which are marked by the dashed lines. 
The twist angle is extracted from the following relation: $\theta\approx\sqrt{\frac{\sqrt{3}n_s}8}a$, where $a$ is the lattice constant of graphene.}
\label{angle_det}
 
\end{figure}

\section{Theoretical density of states (DOS).}

\begin{figure}[H]
\centerline{\includegraphics[width=0.6\textwidth]{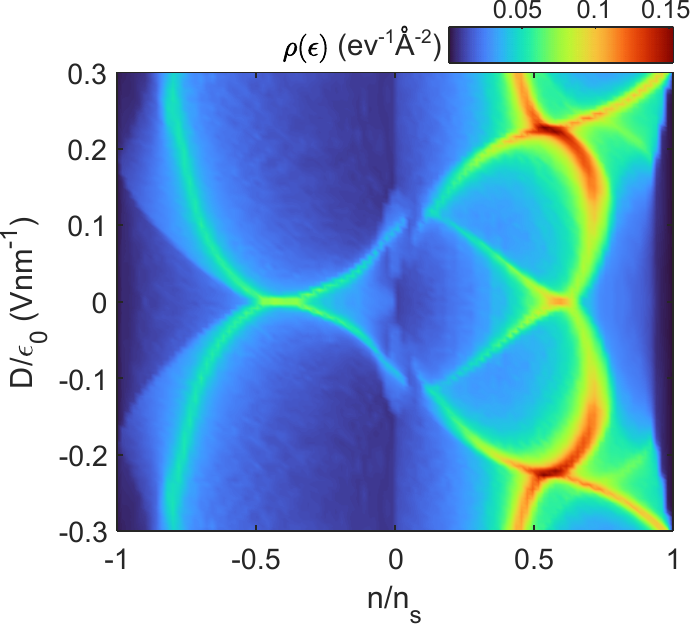}}
\caption{\textbf{Theoretical density of states for device D1 ($\theta\approx1.66^\circ$.)} The theoretical DOS was calculated using the continuum model with the following the ref.\cite{Koshino2019,Mohan2021}. 2D colormap of calculated DOS as a function of $D$ and normalized carrier density. The prominent peaks in the DOS across the $n$-$D$ parameter space identify the locations of van Hove singularities (vHS) and the associated Lifshitz transitions. These features are also reflected in the measured $R_{xx}$ (Fig. 2b) data as weak resistance peaks.}
\label{D1_dos}
 
\end{figure}

\section{Electrical response of D2 device}

\begin{figure}[H]
\centerline{\includegraphics[width=0.6\textwidth]{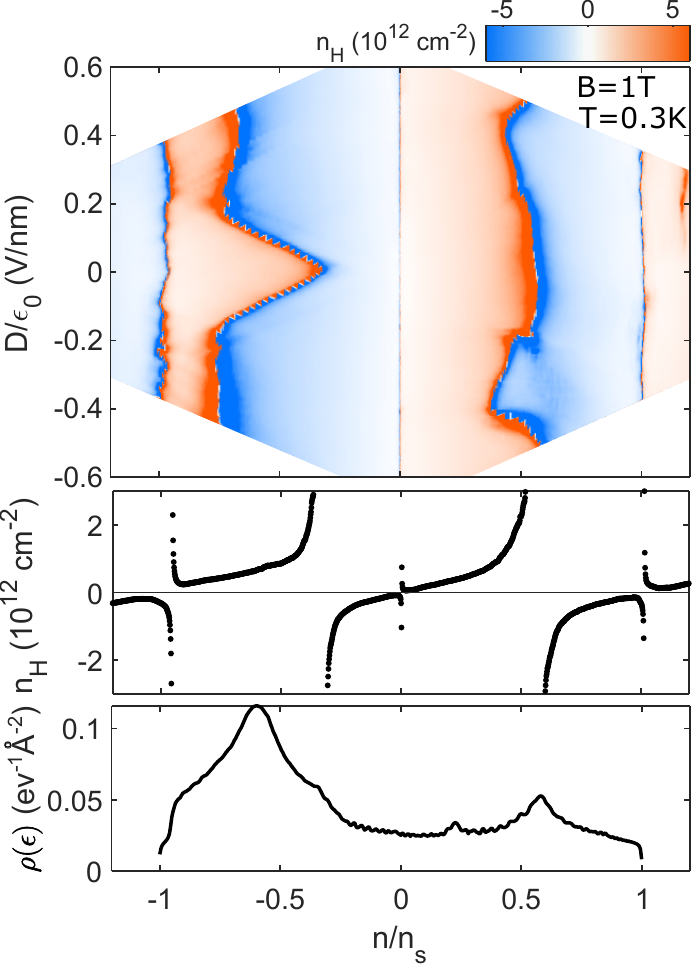}}
\caption{\textbf{Electrical response of device-D2.} (Top panel) Evolution of Hall density, $n_H$ in the $n-D$ parameter space. (Middle) Line cut of $n_H$ at zero $D$. (Bottom) Single particle theoretical DOS at $D=0$.} 
\label{D2_char}
 
\end{figure}

\section{Transverse thermoelectric signal at postive and negative B.}

\begin{figure}[H]
\centerline{\includegraphics[width=0.6\textwidth]{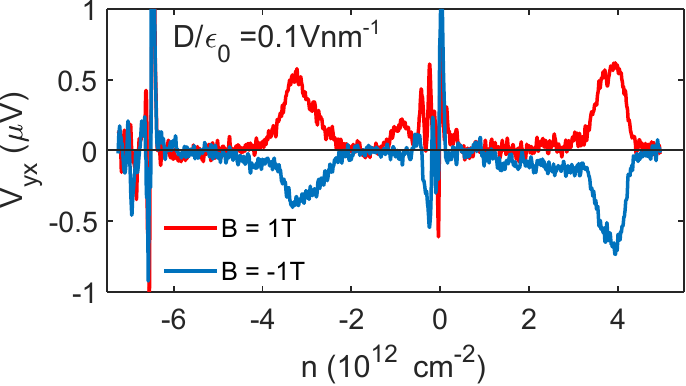}}
\caption{\textbf{Raw data of transverse thermoelectric voltage of device D1.} Thermoelectric voltage  measured with heater excitation of $60\mu A$ and at magnetic field of $B=+1T$ (red) and $B=-1T$ (blue) as function of carrier density for $D/\epsilon_0 = 0.1V nm^{-1}$. Transverse voltage $V_{yx}$, showing a clear sign reversal upon reversing the magnetic field direction. The asymmetry in the signal arises due to finite longitudinal component, because of that all the data plotted in the manuscript was anti-symmetrized.} 
\label{D1_raw_syx_sxx}
\end{figure}

\section{Longitudinal thermopower data.}
\noindent\textbf{Seebeck Coefficient.} We compare the measured $S_{xx}$ [SI-Fig.6 - panel (a)] with the semi-classical formula:
\begin{equation}
    S_{xx}=-\frac{\pi^2}{3}\frac{k_B^2T}{e}(\sigma_{xx}\frac{\partial\sigma_{xx}}{\partial n}+\sigma_{xy}\frac{\partial\sigma_{xy}}{\partial n})\rho(n)
    \label{S_xx1}
\end{equation}
This expression (SI-Fig.6 - panel (c)) correctly reproduces the overall sign change of the dominant carriers across different filling densities $n$ values. 
We also compare with the Mott expression:
\begin{equation}
    S_{xx}=\frac{\pi^2}{3}\frac{k_B^2T}{e}\frac{1}{R_{xx}}\frac{\partial R_{xx}}{\partial n}\rho(n)
    \label{Sxx2}
\end{equation}
which (SI-Fig.6 - panel (b)) 
is in excellent agreement with the experimental data.
\begin{figure}[H]
\centerline{\includegraphics[width=1\textwidth]{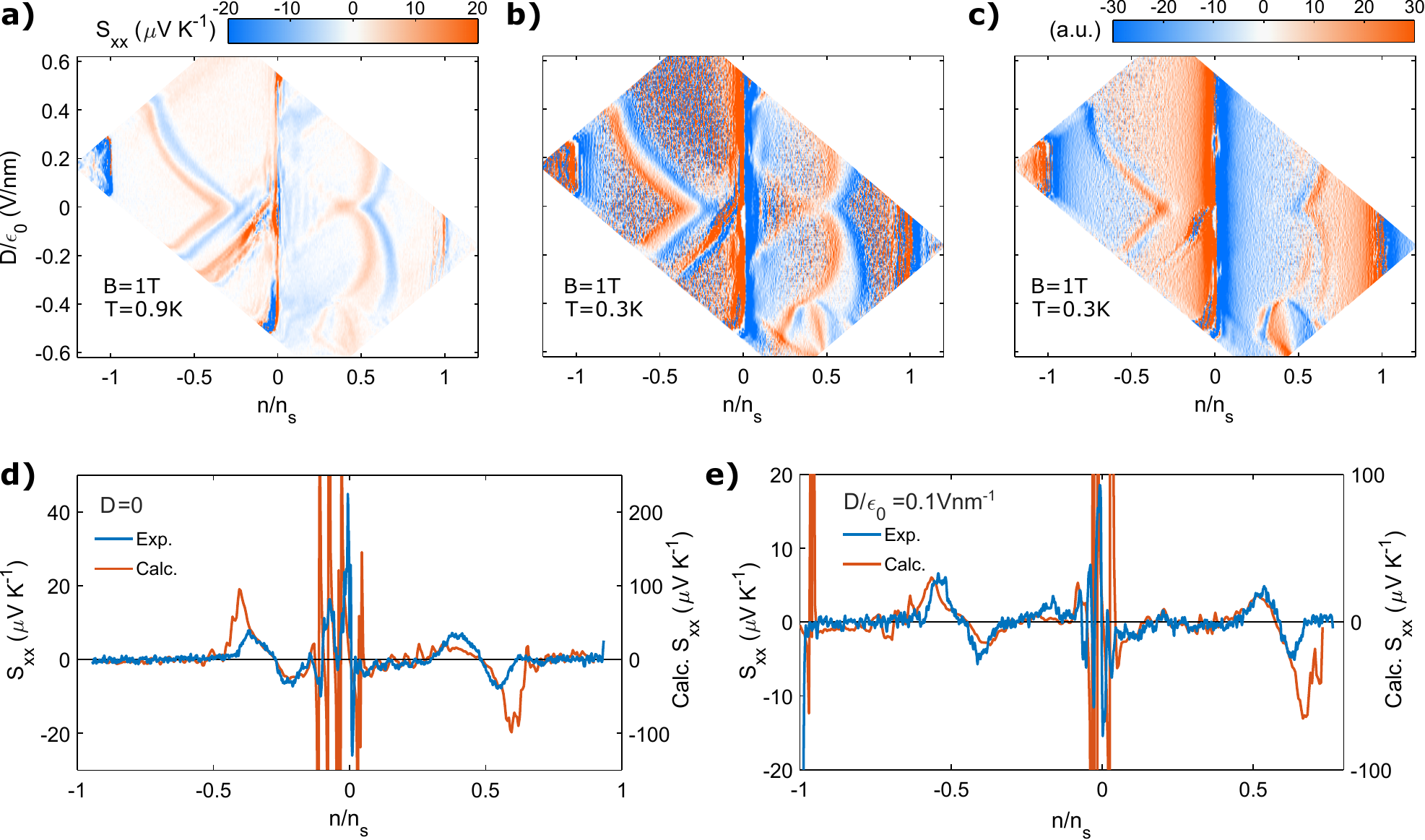}}
\caption{\textbf{Longitudinal thermopower of D1. (a)-(c)} 2D color maps in the $n-D$ plane showing \textbf{(a)} the measured symmetrized $S_{xx}$ at $B = 1 T$ and $T = 0.9K$. \textbf{(b)} The calculated thermopower using \cref{Sxx2}, and (c) the calculated thermopower using \cref{S_xx1}. Note that the DOS contribution, $\rho(n)$ was not included in these calculations. \textbf{(d)-(e)} Direct comparison between measured $S_{xx}$ (orange trace, right axis) and using Mott formula - 
\cref{Sxx2} (blue trace, left axis) at \textbf{(d)} $D = 0$ and \textbf{(e)} $D/\epsilon_0 = 0.1V nm^{-1}$.}
\label{D1_sxx}
 
\end{figure}

\begin{figure}[H]
\centerline{\includegraphics[width=1\textwidth]{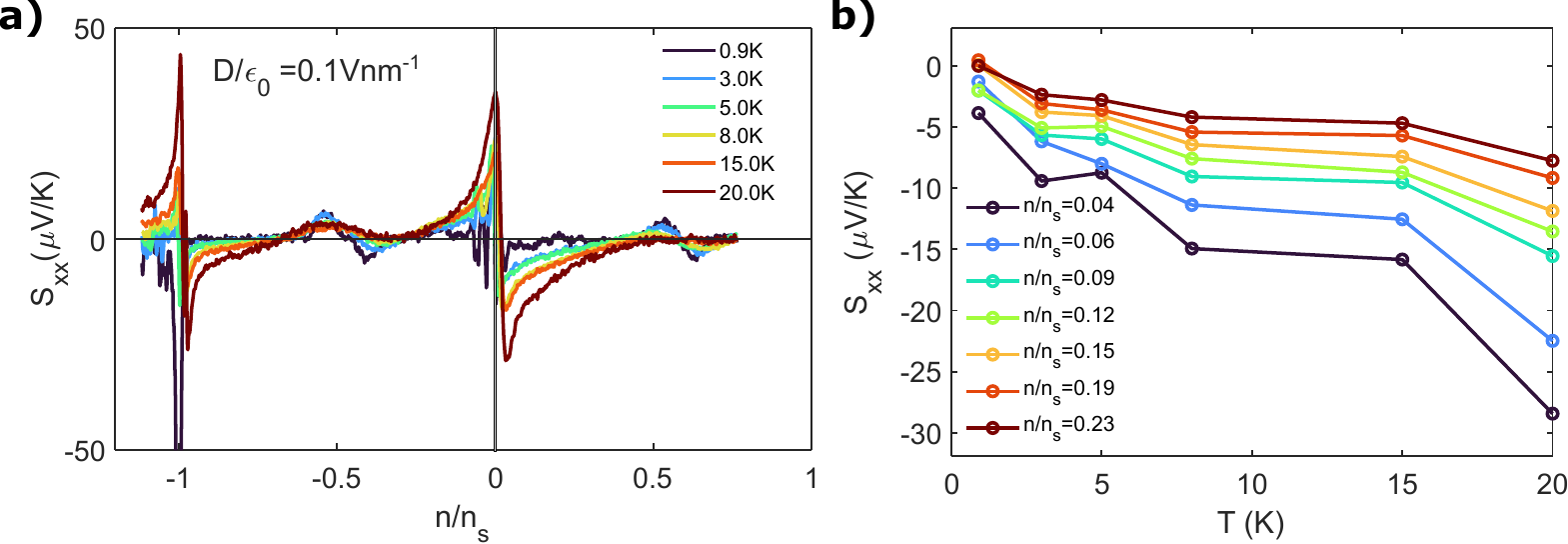}}
\caption{\textbf{Temperature dependent $S_{xx}$ of D1. (a)} $S_{xx}$ as a function of carrier density with increasing temperature at $B=1T$. \textbf{(b)} The cut lines of $S_{xx}$ at different fixed density for the conduction band. The magnitude of the signal increases almost linearly with temperature consistent with the Mott relation.}
\label{d1_sxx_temp}
 
\end{figure}

\begin{figure}[H]
\centerline{\includegraphics[width=1\textwidth]{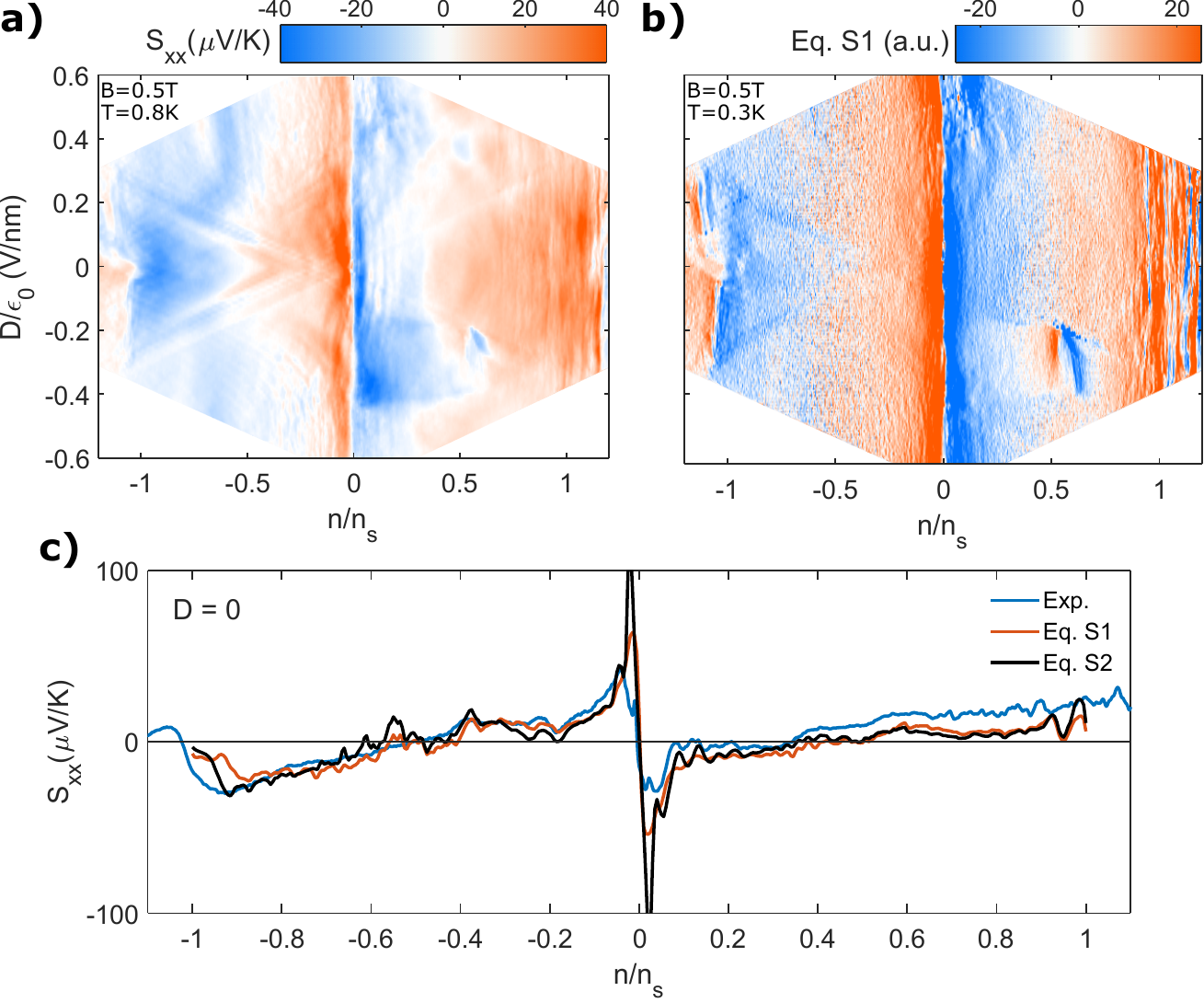}}
\caption{\textbf{Longitudinal thermopower of D2. (a)-(b)} 2D color maps in the $n-D$ plane showing \textbf{(a)} the measured symmetrized $S_{xx}$ at $B = 0.5 T$ and $T = 0.8K$. \textbf{(b)} The calculated thermopower using \cref{S_xx1}, Note that the DOS contribution was not included in these calculations. \textbf{(c)} Direct comparison between measured $S_{xx}$ (blue trace) and the theoretical trace derived from \cref{S_xx1} (red trace) and \cref{Sxx2} (black trace) $D = 0$.}
\label{D2}
 
\end{figure}

 

\section{Transverse thermopower or Nernst signal}

\begin{figure}[H]
\centerline{\includegraphics[width=0.6\textwidth]{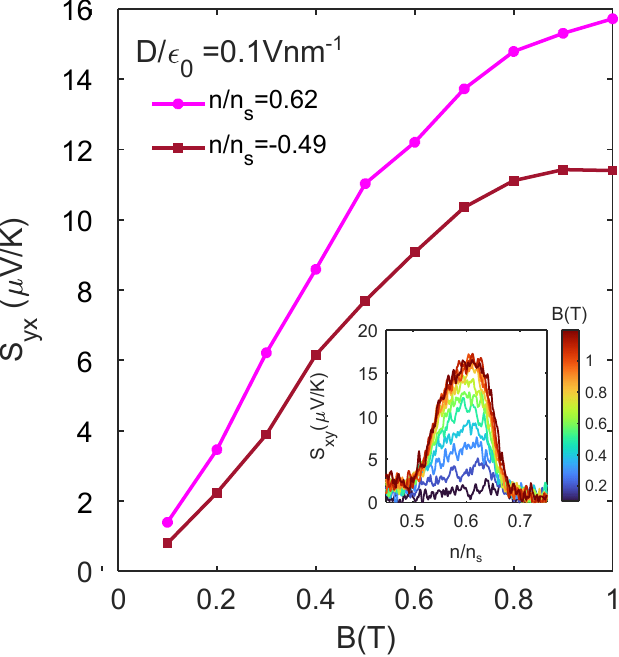}}
\caption{\textbf{Linearity of Nernst signal with $B$ of D1 device.} Inset: $S_{yx}$ 
with $n/n_s$ around the vHS position in the conduction band at $T = 0.9K$. Trace color corresponds to the respective applied magnetic field, as indicated by the color bar. Main: Evolution of $S_{yx}$ for discrete magnetic fields ranging from $100mT$ to $1T$ in increments of $100mT$. Data are plotted for two specific peak positions: $n/n_s = 0.62$ (pink circles) and $n/n_s = -0.49$ (brown rectangles), for the conduction and valency band, respectively. 
The data exhibits a linear dependence on the magnetic field, followed by a saturation of $S_{yx}$ as the field approaches $1T$. Solid lines are provided as a guide for the eye.}
\label{D1_syx_B}
 
\end{figure}

\begin{figure}[H]
\centerline{\includegraphics[width=1\textwidth]{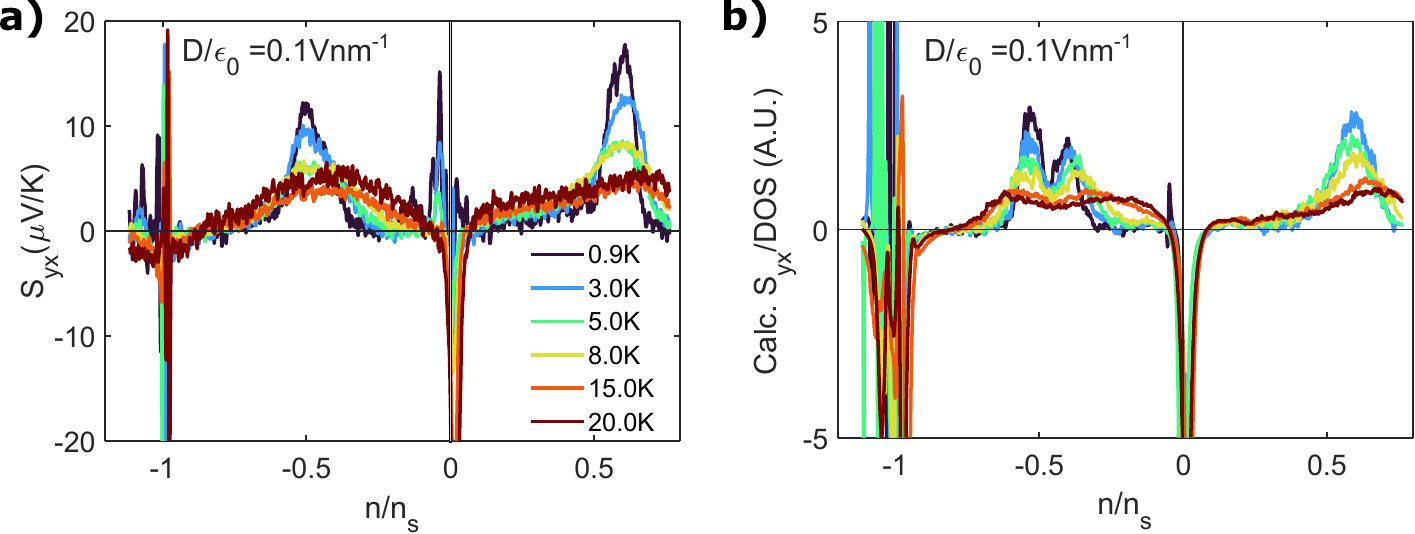}}
\caption{\textbf{Temperature dependence of $S_{yx}$ for D1.} \textbf{(a)} $S_{yx}$ as a function of $n/n_s$ for temperatures ranging from $T=0.9K$ to $20K$ at $B=1T$ and $D/\epsilon_0=0.1Vnm^{-1}$. \textbf{(b)} $S_{yx}$ based on semi-classical equation-Eq.1 of the manuscript and using electrical conductivity tensor. Note that DOS was not included while calculating using Eq. 1 . In both experiment and calculation, the peak amplitude of $S_{yx}$ at the vHS decreases with increasing temperature, confirming that the semiclassical framework captures this thermal suppression.}
\label{D1_syx_tem}
 
\end{figure}

\begin{figure}[H]
\centerline{\includegraphics[width=1\textwidth]{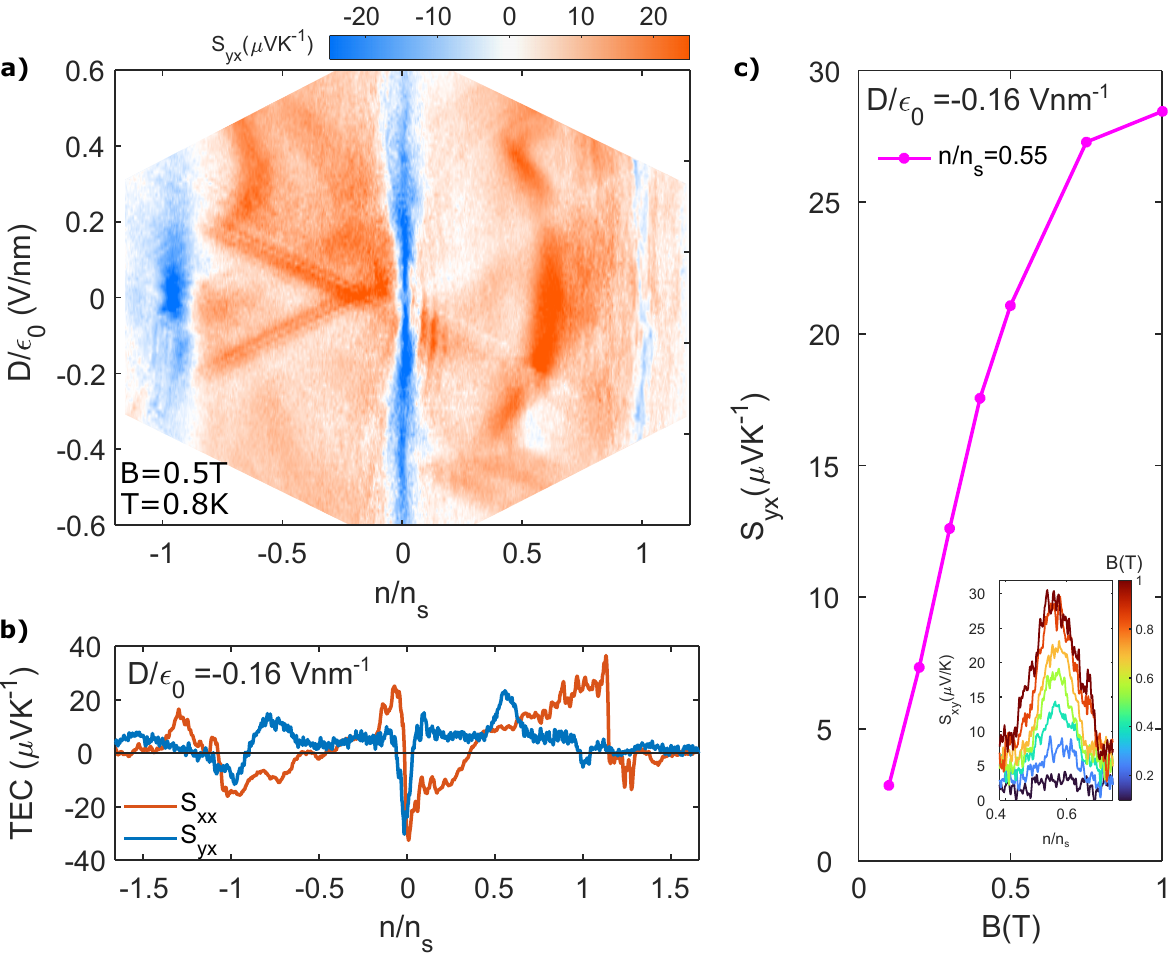}}
\caption{\textbf{Nernst response for D2 device.} \textbf{(a)} 2D colormap of $S_{yx}$. Large postive Nernst signal observed at vHS position as well as negative signal around the Dirac point and moire band fillings. \textbf{(b)} Line cuts at $D=-0.16$ $V nm^{-1}$. \textbf{(c)} Inset shows the $S_{yx}$ with $n/n_s$ in the vicinity of the vHS. The color of each trace corresponds to the applied magnetic field, as indicated by the color bar. Main panel, $S_{yx}$ measured at the peak position (from the inset) vs $B$, exhibiting expected linear response upto $B\approx1T$.}
\label{D2_thermo}
 
\end{figure}
\section{Semiclassical Mott relation}
\begin{figure}[H]
\centerline{\includegraphics[width=1\textwidth]{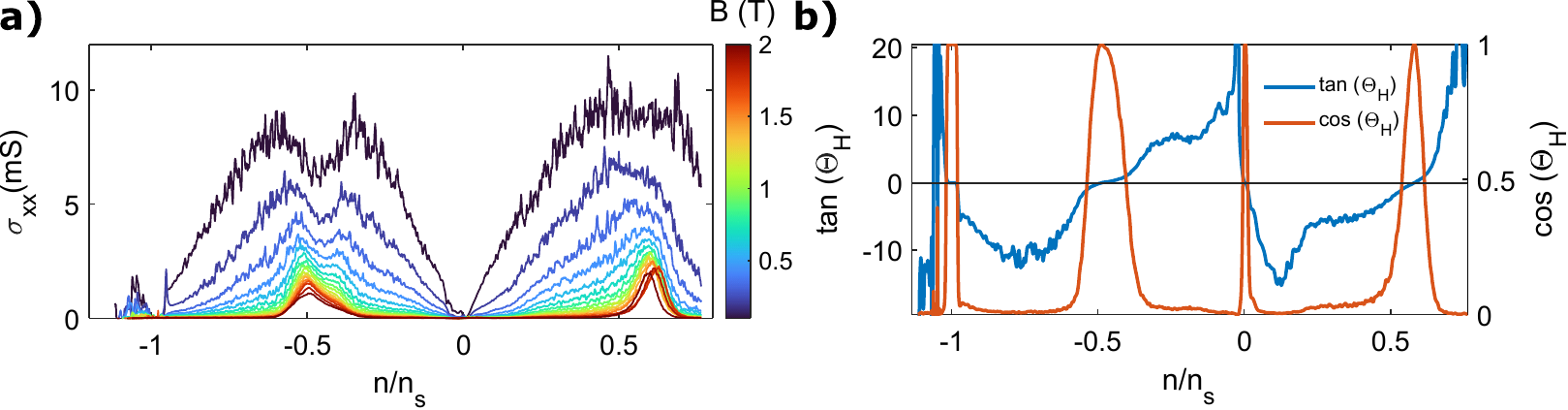}}
\caption{\textbf{D1: Mobility - $\mu$ extraction and Hall angle.} \textbf{(a)} $\sigma_{xx}$ vs $n/n_s$ at $D/\epsilon_0=0.1Vnm^{-1}$. The traces are colored according to the applied magnetic field, ranging from $0.1T$ to $2T$ in increments of $0.1T$ (see color bar). This field-dependent data is used to extract the mobility of the carriers. \textbf{(b)} $\tan(\Theta_H)$ (blue curve, left axis) and $\cos(\Theta_H)$ (orange curve, right axis) as a function of $n/n_s$.}
\label{D1_mu}
 
\end{figure}

\noindent\textbf{Thermoelectric conductivity.} To further demonstrate the consistency between the experimental data and the semiclassical framework, we extracted the thermoelectric conductivity tensor elements ($\alpha_{xx}$, $\alpha_{xy}$) from the experimentally measured longitudinal and transverse transport coefficients ($S_{xx}$, $S_{yx}$, $\sigma_{xx}$, $\sigma_{xy}$) using the following relations:
\begin{equation}
    \alpha_{xx}=\sigma_{xx}S_{xx}+\sigma_{xy}S_{yx}
    \end{equation}
\begin{equation}
    \alpha_{xy}=\sigma_{xy}S_{xx}-\sigma_{xx}S_{yx}
    \end{equation}
\noindent These are then compared with the semiclassical Mott formula: $\bar\alpha=-\frac{\pi^2k_B^2T}{3e}\frac{\partial\bar\sigma}{\partial n}\rho(n)$, where $\rho(n)$ is the DOS.
\begin{figure}[H]
\centerline{\includegraphics[width=1\textwidth]{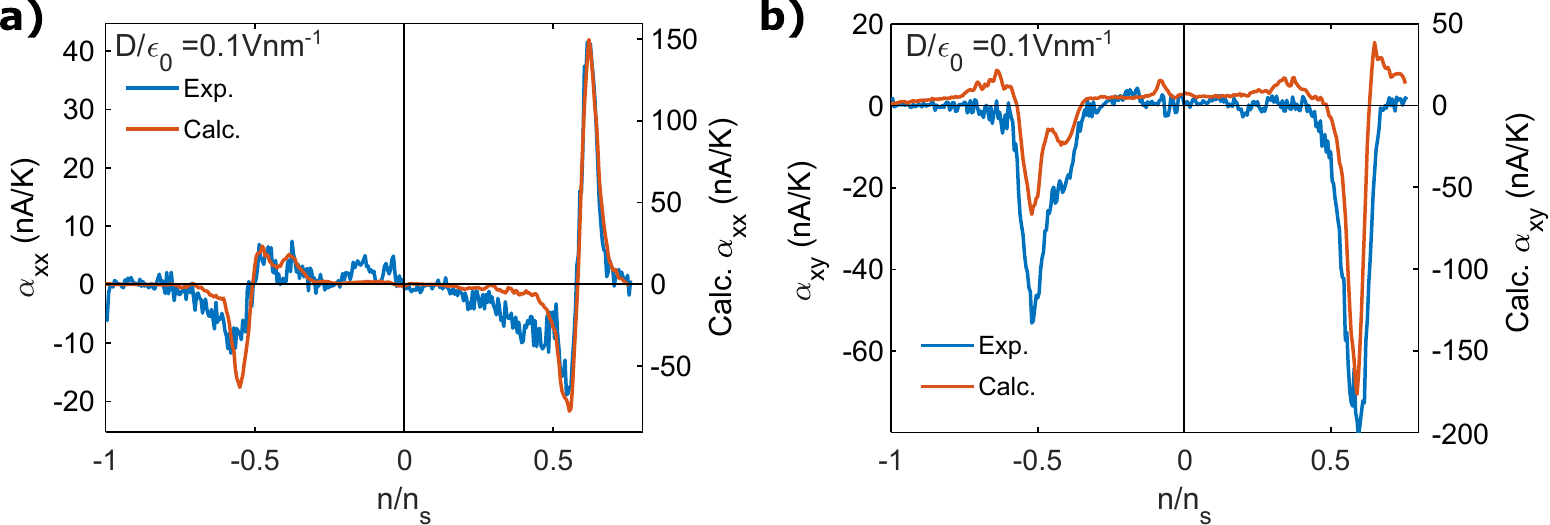}}

\caption{\textbf{Thermoelectric conductivity of D1 device.} Comparison of the thermoelectric conductivity $\alpha$ between experiment and theory as a function of carrier density. The experimental data (blue trace, left axis) is plotted against semiclassical calculations (orange trace, right axis) at a fixed displacement field of $D/\epsilon_0 = 0.1V nm^{-1}$. The experimentally extracted values show excellent agreement with the semiclassical calculations in terms of lineshape and peak positions, with the magnitudes agreeing to within an order of magnitude. (a) Longitudinal component $\alpha_{xx}$ and (b) transverse component $\alpha_{xy}$.}
\label{D1_alpha}
\end{figure}

\begin{figure}[H]
\centerline{\includegraphics[width=1\textwidth]{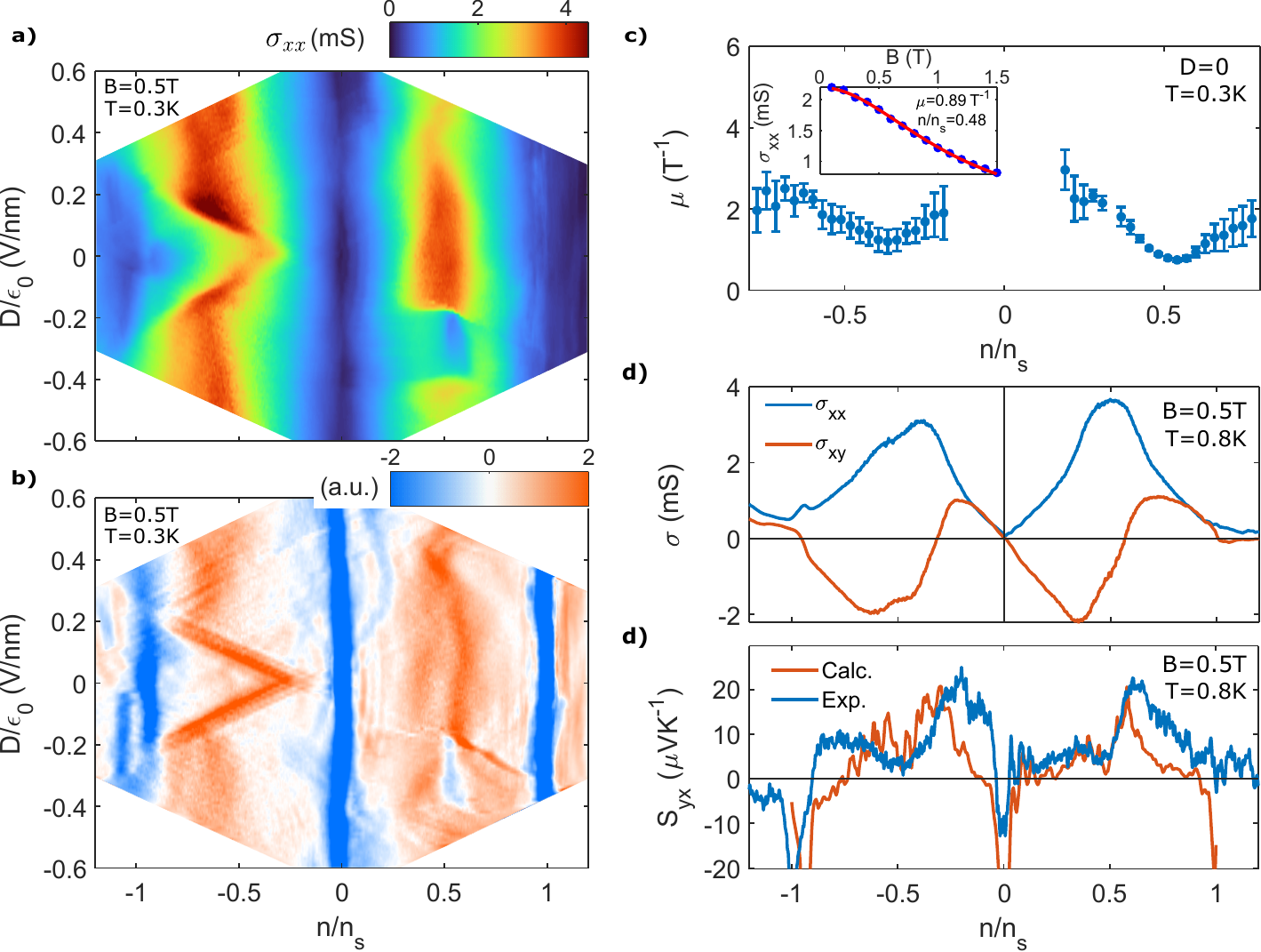}}
\caption{{\textbf{Semiclassical Mott relation of Nernst signal for D2 device. (a)} 2D color map of measured $\sigma_{xx}$. \textbf{(b)} 2D mapping of $S_{yx}$ using Eq. 1 in the same $n-D$ space, without the DOS contribution. \textbf{(c)} Variation of extracted Hall mobility, as a function of carrier concentration at $D=0$. Error bar represent the fitting error during the extraction of $\mu$. Inset shows the magnetic field dependence of $\sigma_{xx}$ (solid circles), at a representative $n/n_s=0.48$ used for the mobility extraction; the red curve is the corresponding fit. The extracted mobility is $0.89T^{-1}$. Due to the large resistive peaks at charge neutrality and full filling, reliable mobility extraction is not possible in those regions. \textbf{(d)} $\sigma_{xx}$ (blue) and $\sigma_{xy}$ vs $B$ at D=0. \textbf{(e)} Comparison between the calculated $S_{yx}$ (Orange trace) obtained using the Mott relation - Eq. 1 with the experimental data (blue trace) from \cref{D2_thermo}a. at $D=0$, plotted as a function of $n/n_s$.}}
\label{D2_analysis}
 
\end{figure}

\section{Seebeck and Nernst signal from a bilayer graphene (BLG) device.}
\begin{figure}[H]
\centerline{\includegraphics[width=1\textwidth]{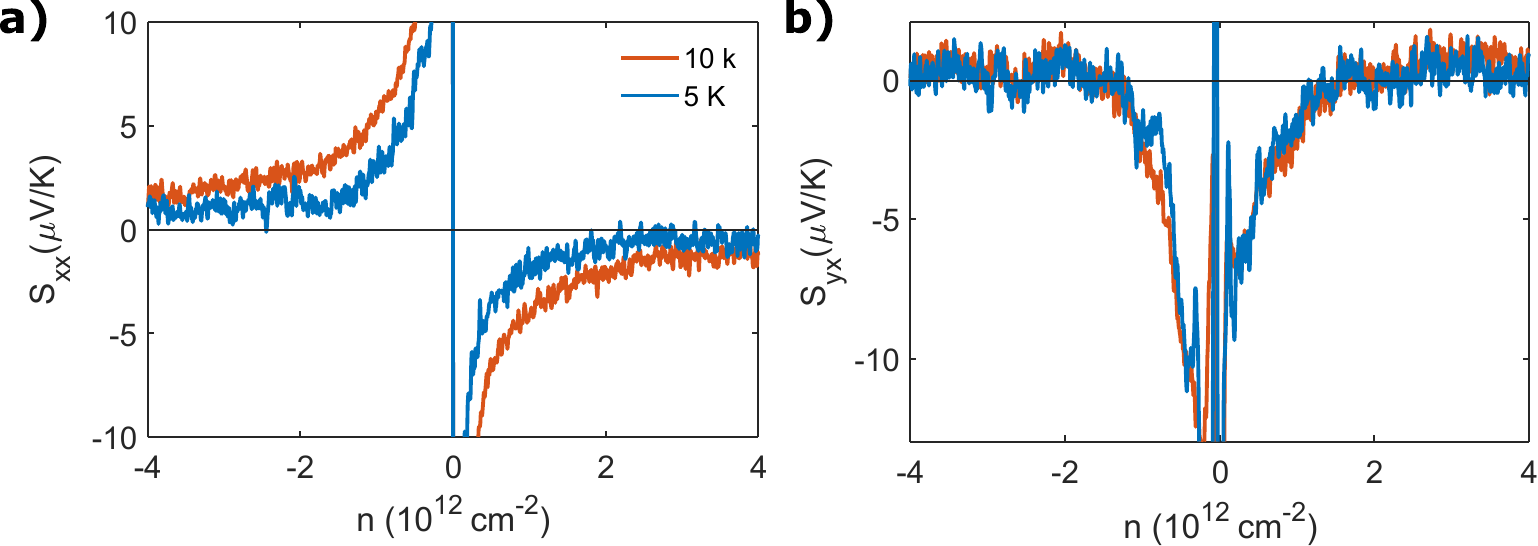}}
\caption{\textbf{$S_{xx}$ and $S_{yx}$ in BLG sample.} Thermopower measured at $B=0.5T$ for temperatures $5K$ (blue trace) and $10K$ (orange trace) as a function on density-$n$. \textbf{(a)} $S_{xx}$ , showing the typical response: positive for hole doping and negative for electron doping, with a sign change at the CNP, pronounced peaks on either side, and a rapid suppression with increasing $|n|$. \textbf{(b)} $S_{yx}$, displays a prominent negative peak at the CNP, which rapidly decreases to zero with increasing carrier density, which is similar to the CNP of the tDBLG as shown in SI-Figure 10. Note that this negative peak around the Dirac point only seen at higher temperature $T \geq 5K$. At lower temperature, the signals are weak, and hard to see any pattern.} 
\label{BLG}
 
\end{figure}

\section{Sign conventions for Hall and thermoelectric measurements}

\begin{figure}[H]

\noindent As we measured both the electrical and thermoelectric transport coefficients, it is important to clarify the sign conventions.

\centerline{\includegraphics[width=0.6\textwidth]{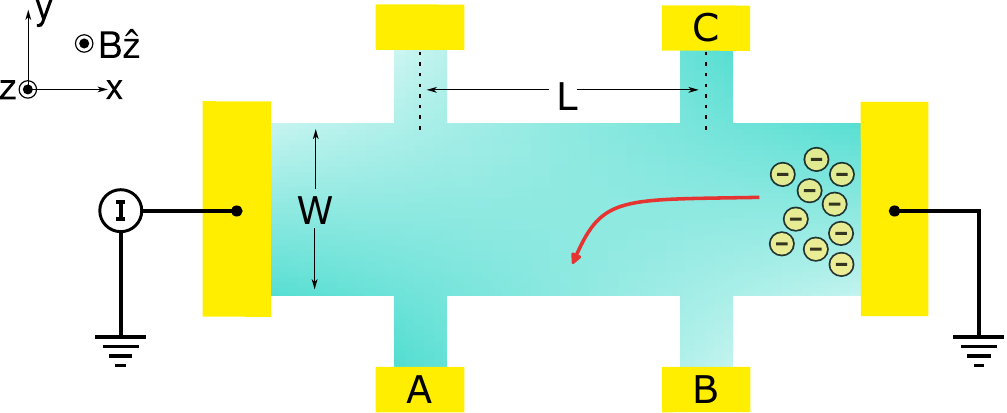}}
\caption{\textbf{Resistance measurements.} Schematics for resistance measurements. The Hall probes are marked by A, B and C. L is the distance between the A and B probes. W is the width of the hall bar. The Hall bar is kept at the external magnetic field $B$ in $+\hat{z}$ direction and biased to current $I$ along the $+\hat{x}$ direction.}
\label{meas_set}
 
\end{figure}

\noindent\textbf{Electrical transport coefficients:} From Drude formula, we can write:
$$\vec{E}=\stackrel{\leftrightarrow}{\rho}\vec{J}$$
where $\vec{E}(E_x, E_y)$ is the electric field, $\stackrel{\leftrightarrow}{\rho}$ is the resistivity tensor and $\vec{J}(J_x,0)$ is the applied current density, as per the axes defined in SI Fig.1. Now,
\begin{align*}
   \rho_{xx}&=\frac{E_x}{J_x}; & \rho_{xy}&=-\frac{E_y}{J_x} \\
   &=\frac{-\nabla_xV}{J_x}; & &=-\frac{-\nabla_yV}{J_x} \\
   &=\frac{-(V_A-V_B)/(x_A-x_B)}{J_x}; & &=-\frac{-(V_B-V_C)/(y_B-y_C)}{J_x} \\
   &=\frac{-(V_A-V_B)/(-L)}{I/W}; & &=-\frac{-(V_B-V_C)/(-W)}{I/W} \\
\end{align*}
Therefore,
\begin{align}
\rho_{xx}&=\frac{W}{L}\frac{V_A-V_B}{I}; & \rho_{xy}=-\frac{V_B-V_C}{I}
\end{align}

\noindent Now, for the given setup in SI Fig.1, let us check the sign for the Hall resistance. With the applied bias, the electrons tend to move in $-\hat{x}$ direction but in the presence the magnetic field, they will deflect in $-\hat{y}$, due to the Lorentz force. Hence, according to S1, the measured $\rho_{xy}$ would be +ve for electronic carriers. Similarly, for hole carriers $\rho_{xy}$ is -ve.

\noindent\textbf{Thermoelectric coefficients:} Unlike Hall or Seebeck, the Nernst coefficient does not depend on the carrier type. So in modern literature the vortex convention is currently used to determine the polarity of the Nernst coefficient. Here, also the same convention has been used.

\noindent In order to measure the thermoelectric voltages (TEV), we employ the traditional $2\omega$ method. A sinusoidal current, $\tilde{I}$ of frequency $\omega$ is applied to the heater (SI-Figure 17), in response the temperature gradient, $\propto\tilde{I}^2$ is generated. 
Now,
$$TEV\propto sin^2(\omega t)\propto -cos(2\omega t)$$
Therefore, the TEV generated will oscillate with twice the frequency and will be in $-90^\circ$ phase difference to the excitations applied to the heater. In practice, lock-in amplifiers are used to measure the TEV, where the 2nd harmonics in $-90^\circ$ phase difference to applied $\tilde{I}$, are captured. The extra `-ve' sign needs to accounted while converting the thermoelectric coefficients from measured TEV.

\begin{figure}[H]
\centerline{\includegraphics[width=0.5\textwidth]{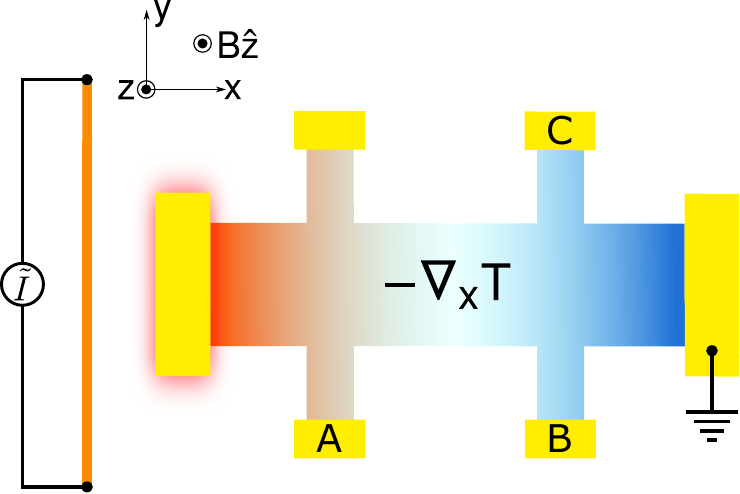}}
\caption{\textbf{Thermoelectric measurements.} Schematics for the thermoelectric measurements. A temperature gradient, $-\nabla T$ in $\hat{x}$ direction is created by applying a current $\tilde{I}$ to the microscopic heater near to the device. The longitudinal (Seebeck) and transverse (Nernst) thermoelectric voltages formed in response, are measured between hall probes $A-B$ and $B-C$.}
\label{meas_set2}
 
\end{figure}

\noindent For the thermal transport, we can write:
$$\vec{E}^{th}=\stackrel{\leftrightarrow}{S}\vec{\nabla}T$$
Here, $\vec\nabla T(\nabla_xT,0)$ is the applied temperature gradient, $\stackrel{\leftrightarrow}{S}$ is the thermoelectric tensor, and $\vec{E}^{th}(E^{th}_x, E^{th}_y)$ is the generated electric field (in accordance with vortex convention\cite{Behnia2009}) due to thermal flow of carriers. The longitudinal and transverse thermoelectric coefficients $S(S_{xx})$ (Seebeck) and $N(S_{xy})$ (Nernst) defined as,
\begin{align*}
    S_{xx}&=\frac{E^{th}_x}{\nabla_xT}; & S_{yx}&=\frac{E^{th}_y}{\nabla_xT} \\
    &=\frac{-\nabla_xV}{\nabla_xT}; & &=\frac{-\nabla_yV}{\nabla_xT} \\
    &=\frac{-(V_A-V_B)/(x_A-x_B)}{(T_A-T_B)/(x_A-x_B)}; & &=\frac{-(V_B-V_C)/(y_B-y_C)}{(T_A-T_B)/(x_A-x_B)} \\
    &=-\frac{V_A-V_B}{T_A-T_B}; & &=\frac{-(V_B-V_C)/-W}{(T_A-T_B)/-L}
\end{align*}
Hence,
\begin{align}
    S_{xx}&=-\frac{V_A-V_B}{\Delta T}; & S_{yx}&=-\frac{V_B-V_C}{\Delta T}\frac{L}{W}
\end{align}

\noindent It can be seen from SI Figure 17 that, for electrons, $V_A-V_B$ is positive, so $S_{xx}$ is negative. Similarly, for holes, it will be vice-versa. Therefore, the sign for longitudinal thermopower is unique. However, for transverse thermopower, $S_{yx}$, as discussed in the manuscript and shown in SI Figure 18, the signs are not unique except at the vHSs and Dirac point. In the conduction or valence band (as marked by $E_{F1}$ and $E^{'}_{F1}$), the competition between the hot and cold carriers deflected to the opposite directions due to the magnetic field, and will decide the sign of the $V_B-V_C$ in Eq. S6, and thus the sign of the $S_{yx}$ is not unique. In contrast, at vHS (as marked by $E_{F2}$ and $E^{'}_{F2}$), the $V_B-V_C$ is negative, and thus $S_{yx}$ will be positive. Note that the deflection for the hot carriers is different around the vHS due to the change of the sign of the mass as discussed in the manuscript. Similarly, around the Dirac point ($E_{F0}$), the $V_B-V_C$ is positive (as shown in SI Figure 18 (b)-bottom-panel), and thus $S_{yx}$ will be negative.


\begin{figure}[H]
\centerline{\includegraphics[width=1\textwidth]{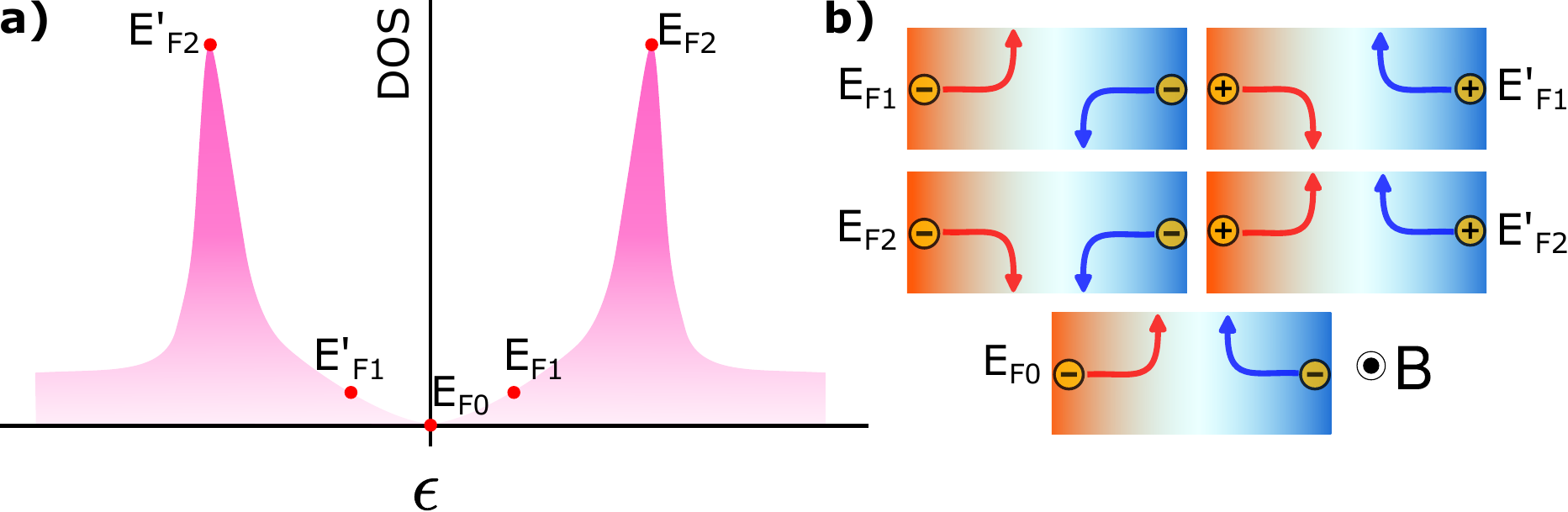}}
\caption{\textbf{The sign of Nernst signals in band. (a)} A schematic representation of DOS of twisted system, where the large DOS indicates the position of the vHSs in the conduction and valence bands.  \textbf{(b)} The trajectory of hot (above the chemical potential from the hot end, represented by the red line) and cold (below the chemical potential from the cold end, represented by the blue line) carriers in the presence of a magnetic field. This presentation is done considering one type of carriers for the conduction band --- electrons, and for the valence band --- holes. As can be seen for the conduction or valence band (as marked by $E_{F1}$ and $E^{'}_{F1}$), the competition between the hot and cold carriers deflected to the opposite directions in the presence of a magnetic field, and the sign will be determined by the resultant voltage along the transverse direction, and thus the sign of the $S_{yx}$ is not unique. In contrast, at vHS (as marked by $E_{F2}$ and $E^{'}_{F2}$), the transverse voltage ($V_B-V_C$) will be always negative, and thus $S_{yx}$ will be positive. Note that the deflection for the hot carriers is different around the vHS due to the change of the sign of the mass as discussed in the manuscript. Around the Dirac point ($E_{F0}$), the $V_B-V_C$ is positive and thus $S_{yx}$ will be negative. Note that around the Dirac point, the cold electrons are from the valence band, and thus have negative mass, in contrast to hot electrons arising from the conduction band; as a result, they bend to the same side. Around the Dirac point, we observe negative $S_{yx}$ only at higher temperatures, whereas large positive $S_{yx}$ was measured at vHSs at lower temperatures. This is the importance of vHS, even though the $k_BT$ window is smaller, but due to the large magnitude of $J_H$ and $J_C$ due to enhanced DOS (as shown in Fig. 1d of the manuscript), they produce large $S_{yx}$.}
\label{carr_dyn}
\end{figure}
\section{Estimation of temperature gradient.}
\noindent In order to estimate the temperature gradient we utilize multiterminal resistance measured as a function of both heater current and bath temperature, as illustrated in the schematic below. First, the longitudinal resistance between adjacent Hall-bar contacts along the length of the sample is recorded as a function of the applied DC heater current ($I_H$) while keeping the bath temperature fixed. Subsequently, the heater current is set to zero, and the bath temperature is varied while measuring the resistance between the same Hall-bar pairs. The resistance variations obtained from these two measurements are then compared to establish a relationship between $I_H$ and the corresponding temperature inferred from the resistance values. This temperature can be interpreted as the local temperature ($T_{Loc}$) at the midpoint of a given Hall-bar pair. This procedure yields a $T_{Loc}-I_H$ curve for each Hall-bar pair along the direction of the applied temperature gradient. The temperature difference $\Delta T$ between two successive Hall-bar pairs is then obtained by subtracting their respective $T_{Loc}$ values, and dividing $\Delta T$ by the physical separation between the midpoints of the two pairs gives the local temperature gradient.
\begin{figure}[H]
\centerline{\includegraphics[width=0.5\textwidth]{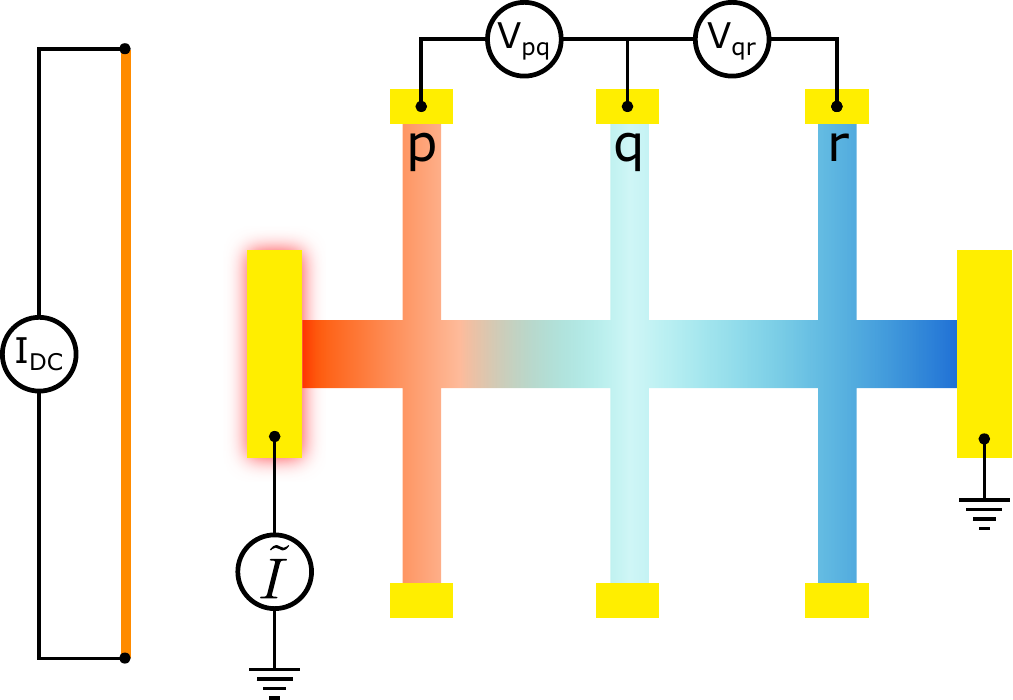}}
\caption{\textbf{Schematic.} A direct current ($I_H$) is passed through an isolated micro-heater positioned adjacent to the device, generating a steady-state temperature gradient along the length of the sample. The four-terminal resistance ($R=\frac{V}{\tilde{I}}$) is then measured between successive Hall-bar contact pairs p-q and q-r, which are positioned along the direction of the heater-induced temperature gradient.}
\label{temp_cal_sch}
 
\end{figure}

\begin{figure}[H]
\centerline{\includegraphics[width=1\textwidth]{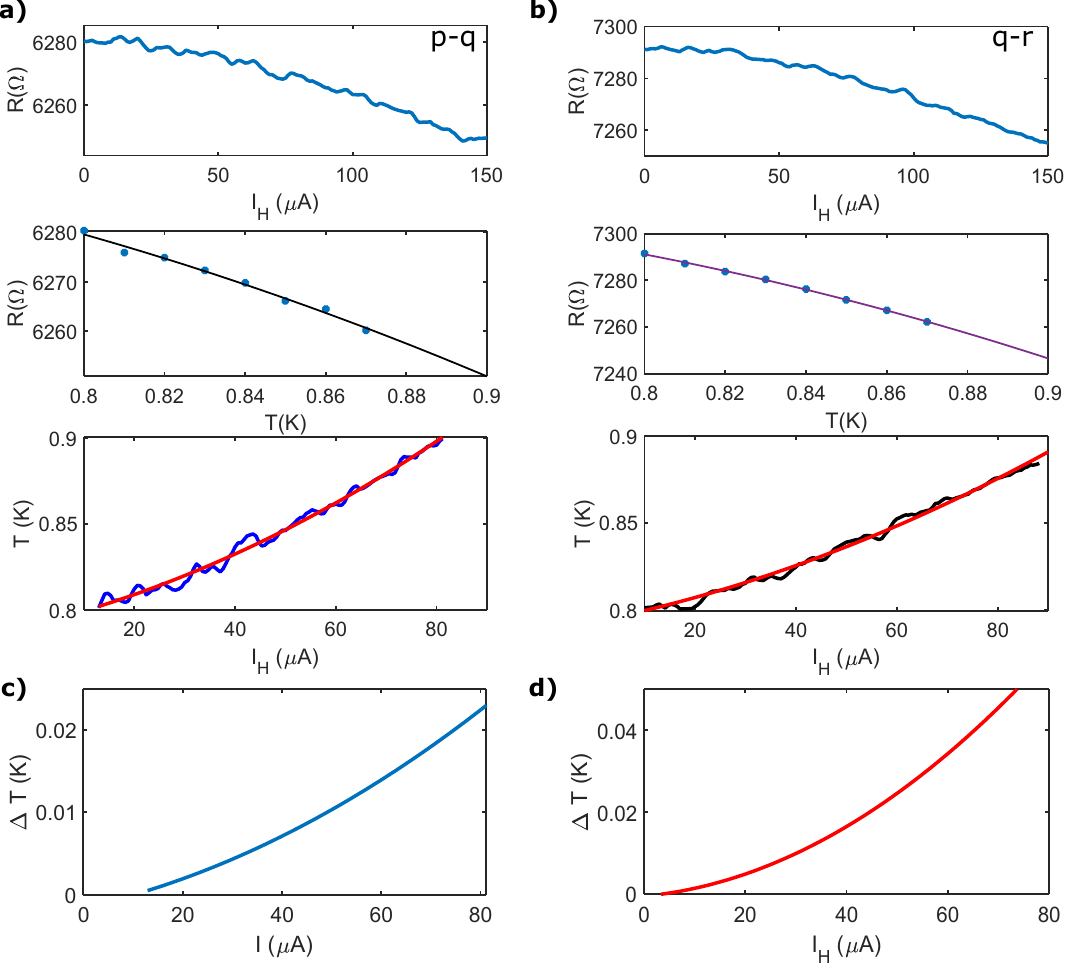}}
\caption{\textbf{Data used for the estimation of the temperature gradient. (a)-(b)} From top to bottom: longitudinal resistance $R$ as a function of heater current $I_H$; $R$ as a function of bath temperature $T_{bath}$. Solid circles are the data points and solid line is the fitted curve extrapolated upto $0.9$ $K$; extracted local temperature $T_{Loc}$ vs $I_H$ (red curve denotes the fit), obtained from contact pairs p-q \textbf{(a)} and from q-r \textbf{(b)} measured in device D2. \textbf{(c)} $\Delta T$ vs $I_H$ derived from (a) and (b) at bath temperature $\sim 0.8K$. \textbf{(d)} $\Delta T$ vs $I_H$ for device D1 obtained in similar way.}
\label{D2_}
 
\end{figure}

\section{Minimal model for positive Nernst peak due to Van-Hove Singularity \label{MIN_MODEL_ARKA}}
\begin{figure}[htbp]
\centerline{\includegraphics[width=1\textwidth]{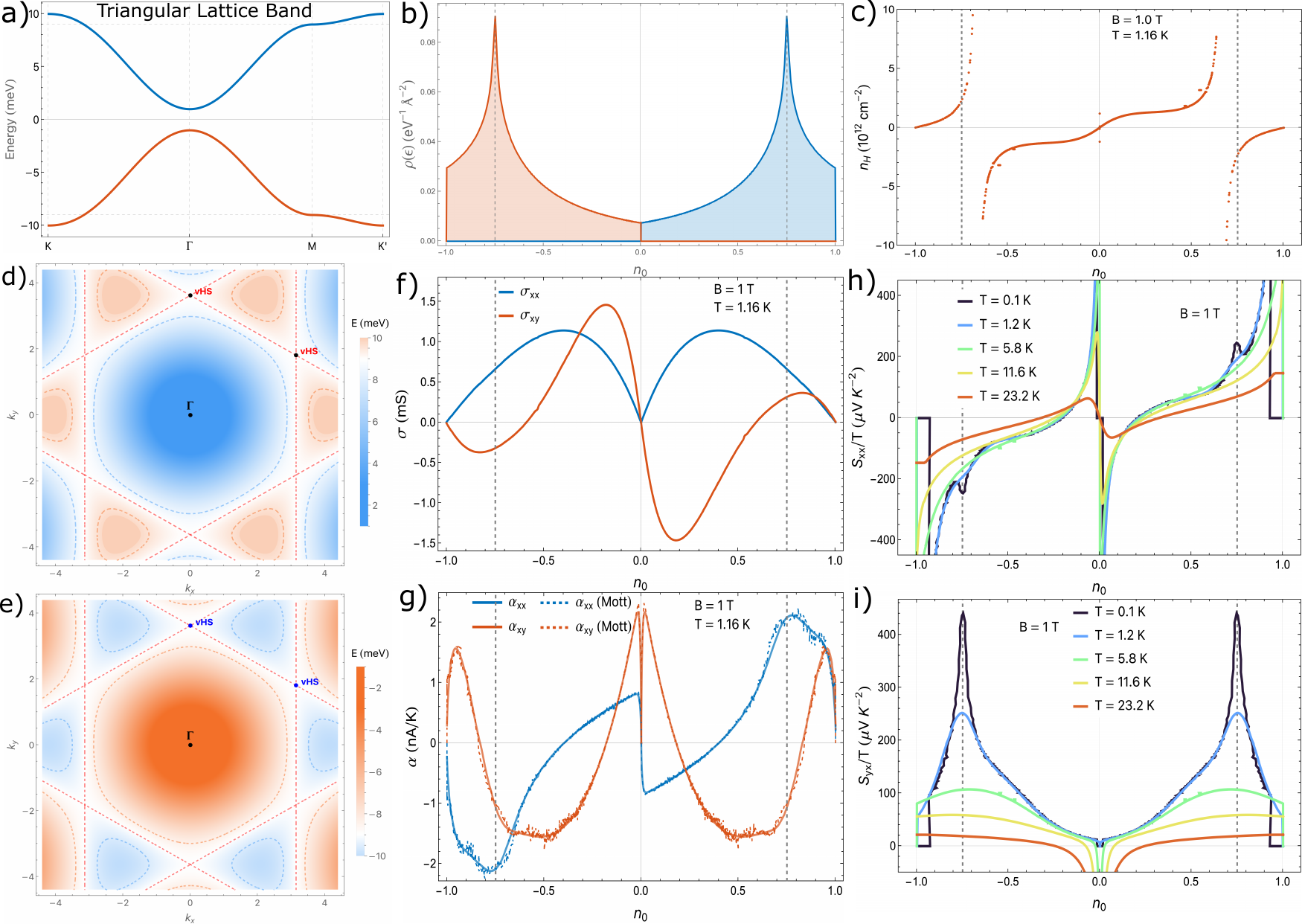}}
\caption{
\textbf{Band structure and transport coefficients for a minimal two-band model.} \textbf{(a)} Triangular lattice tight-binding model (eq.~\ref{min_triag_mod}) with nearest-neighbor interactions ($t_e=t_h=1$), and offsets $\varepsilon^{0}_{e}=-\varepsilon^{0}_{h}=1$. 
\textbf{(b)} Density of states \textbf{(c)} and Hall density as functions of the filling fraction ($n_{0}$). DOS diverges at $n_{\text{vHS}}=\pm 0.75$ (denoted by a black dotted line). \textbf{(d), (e)} Fermi surface contours below (orange), at (red), and above (blue) the vHS for each bands. \textbf{(f)} $\sigma_{xx}$ (blue) and $\sigma_{xy}$ (orange); \textbf{(g)} $\alpha_{xx}$ (solid blue) and $\alpha_{xy}$ (solid orange) evaluated using the exact formalism (eq.~\ref{sig-alph-arka}) compared with the Mott relation (dotted lines) showing extremely good agreement.  \textbf{(h)} Seebeck coefficient ($S_{xx}$) and \textbf{(i)} Nernst coefficient ($S_{yx}$) (Eq.~\ref{see-arka}) for different temperatures as a function of $n_{0}$. Nernst signal peaks exactly at $n_{\text{vHS}}$
}
\label{fig_arka_SI_3}
\end{figure}

Here, we consider a minimal model comprising two non-interacting, tight-binding triangular lattice bands. We assume equal hopping parameters ($t_{e}=t_{h}=1$) for both bands, separated by an energy gap of $\varepsilon^{0}_{e}-\varepsilon^{0}_{h}$. The dispersions for the electron ($\alpha=e$) and hole ($\alpha=h$) bands are defined with their corresponding signs in Eq.~\ref{min_triag_mod}. Both of these bands are illustrated in SI-Fig.~\ref{fig_arka_SI_3}(a). For each band, we identify the van Hove singularities (vHS) at $\mathcal{E}_{\text{vHS}} = \pm 9 t_{e}$ (located at the $M$ points of the Brillouin zone). The density of states for electrons and holes can then be calculated using Eq.~\ref{DOS-arka}:
\begin{eqnarray}
\displaystyle
\mathcal{E}_{\alpha}(k) = \varepsilon^{0}_{\alpha}\pm 6t_{\alpha}\pm t_{\alpha}\varepsilon(k) ,\quad
\displaystyle
\varepsilon(k) = -2 \left[ \cos(k_{x}) + 2 \cos(k_{x}/2) \cos(\sqrt{3} k_{y}/2) \right]
\label{min_triag_mod}
\\
\displaystyle
D_{\alpha}(\mathcal{E}_{\alpha}) = \frac{\sqrt{3/2}}{2\pi^{2}\sigma \sqrt{\pi}}\int_{-\frac{\pi}{2}}^{\frac{\pi}{2}}\int_{-\frac{\pi}{\sqrt{3}}}^{\frac{\pi}{\sqrt{3}}}d\theta_{x}d\theta_{y} \exp \left[ -\frac{[\mathcal{E}-\mathcal{E}_{\alpha}(\theta)]^2}{2\sigma^2} \right]
\label{DOS-arka}
\end{eqnarray}
The distributions of electrons and holes are governed by the corresponding Fermi-Dirac functions: $\displaystyle f_{e}(\mathcal{E}_{e})=\left[\exp\left[\frac{\mathcal{E}_{e} -\mu}{k_{B}T}\right]+1\right]^{-1}$, $\displaystyle f_{h}(\mathcal{E}_h)=1-f_{e}(\mathcal{E}_h)$. Here, $\mu$ is the chemical potential at temperature $T$. To ensure particle conservation, $\mu$ must vary with $T$ such that the net carrier density, $n_{\text{net}}(T)=n_{e}(T)-n_{h}(T)$, remains fixed at its zero-temperature value, $n_{0}$. 
In SI-Fig.~\ref{fig_arka_SI_3}(c), we plot the 
density of states (DOS) as a function of the net carrier density ($n_0$). Here, we observe a logarithmic divergence in the DOS at $n_{0} = \pm 0.75$, which is a hallmark of a vHS in two dimensions. In SI-Figs.~\ref{fig_arka_SI_3}(d) and (e), we illustrate the Fermi surface contours (dotted lines) for both bands at energies below the vHS (orange), exactly at the vHS (red), and above the vHS (blue). At vHS point the Fermi surface curvature changes showing a Lifshitz transition.
\\
Building on the standard Boltzmann transport equation, we define the specific transport quantities for each band. We express the electric current in terms of an applied electric field ($\vec{E}$) and a thermal gradient ($\nabla T$), which are governed by the electrical conductivity tensor ($\sigma$) and the thermoelectric conductivity tensor ($\alpha$).
\begin{equation}
\mathbf{J} = \boldsymbol{\sigma} \vec{E} + \boldsymbol{\alpha} \vec{\nabla}(T)
\end{equation}
For two-dimensional system with a magnetic field applied along the $z$-direction, $\vec{B}=B_{z}\hat{z}$ transport functions  significantly simplifies, yielding diagonal and off-diagonal components defined as:
\begin{eqnarray}
\displaystyle
\sigma_{xx} = \frac{e^2}{\hbar V} \sum_k \tau \left( -\frac{\partial f_0}{\partial \varepsilon} \right) v_x v_x;
&
\displaystyle
\alpha_{xx}= \frac{-e k_B}{\hbar V} \sum_k \tau \left( -\frac{\partial f_0}{\partial \varepsilon} \right) v_x v_x(x-\tilde{\mu})
\\
\displaystyle
\quad\quad \sigma_{xy}=- \frac{e^3}{\hbar V} \sum_k \tau^{2} \left( -\frac{\partial f_0}{\partial \varepsilon} \right) v_x v_x \frac{\partial v_y}{\partial k_y}  B_{z};
&
\displaystyle
\alpha_{xy}=\frac{e^{2} k_B }{\hbar V} \sum_k \tau^{2} \left[ -\frac{\partial f_0}{\partial \varepsilon} \right] \frac{(\epsilon - \mu)}{T}v_x^{2} B_z \frac{\partial v_y}{\partial k_y}~~~~~~~~
\label{sig-alph-arka}
\end{eqnarray}
Here $\tau$ is the relaxation time, which for all of our current calculation is set to a constant value $\tau=\tau_{0}=1$. 
\newline
For defining Nernst and Seebeck coefficient we set to an open circuit configuration $J_x=J_y=0$ condition with thermal gradient applied only in $x$ direction ($\nabla_y T =0$)
\begin{equation}
\displaystyle
\displaystyle S_{yx}=\frac{E_y}{\nabla_x T} = \frac{\sigma_{xx}\alpha_{yx}-\sigma_{yx}\alpha_{xx}}{\sigma_{xx}\sigma_{yy}-\sigma_{yx}\sigma_{xy}}
\quad\quad
\displaystyle S_{xx}=\frac{E_x}{\nabla_x T} = \frac{\sigma_{yy}\alpha_{xx}-\sigma_{yx}\alpha_{xy}}{\sigma_{xx}\sigma_{yy}-\sigma_{yx}\sigma_{xy}}
\label{see--arka}
\end{equation}
Following this approach, we compute each of these conductivities for both bands. The total macroscopic transport quantities are obtained by summing the individual band components, carefully accounting for the sign of the charge carriers. Specifically, the Hall conductivity ($\sigma_{xy}$) and the longitudinal thermoelectric conductivity ($\alpha_{xx}$) change sign depending on the carrier type, whereas the contributions to the remaining two quantities ($\sigma_{xx}$ and $\alpha_{xy}$) are strictly additive.  We plot these total conductivities in SI-\textbf{Figs.~\ref {fig_arka_SI_3} \textbf{(f)} and \textbf{(g)}}, and subsequently use them to calculate the overall Seebeck and Nernst coefficients. It is important to emphasize that, because they are derived via tensor inversion, the total Seebeck and Nernst signals are not simply the direct sums of their individual single-band counterparts.
\newline
In SI-Fig.~\ref{fig_arka_SI_3}(c), we plot the Hall density, $n_H = -B_z \Delta^2/(e \sigma_{xy})$ (where $\Delta^{2}=\sigma_{xx}^{2}+ \sigma_{xy}^2$). $n_H$ changes sign at the charge neutrality point (CNP) due to the carrier switch. Near the vHS, $n_H$ exhibits a Lifshitz-driven divergence that is slightly offset from the vHS energy, matching the $\sigma_{xy}$ zero-crossings in Fig.~\ref{fig_arka_SI_3}(f).
\newline
In SI-Fig.~\ref{fig_arka_SI_3}(f), $\sigma_{xx}$ remains strictly positive and exhibits a peak slightly offset from the vHS. Its steepest slope occurs exactly at the vHS ($n_{\text{vHS}}=\pm 0.75$). Conversely, $\sigma_{xy}$ changes sign at the charge neutrality point (CNP) and exhibits additional zero-crossings offset from the vHS. In SI-Fig.~\ref{fig_arka_SI_3}(g), $\alpha_{xx}$ exhibits a prominent peak near the vHS, while $\alpha_{xy}$ undergoes a distinct sign change. These features are direct consequences of the semiclassical Mott relation, the predictions of which are plotted as dotted lines in Fig.~\ref{fig_arka_SI_3}(g).The origin of this offset can be explained by the definition of electrical conductivity, which depends on both the density of states and the carrier velocity squared ($\sigma_{xx} \propto \int d\mathcal{E} D(\mathcal{E})v_{x}^{2}$). At a vHS, the density of states diverges logarithmically, whereas the velocity squared approaches zero polynomially. Due to these competing effects, $\sigma_{xx}$ peaks near the vHS, rather than exactly at it. Similarly for $\sigma_{xy}$, this competition shifts the exact cancellation of transport contributions from both bands away from the singularities.
\newline
In SI-\textbf{Figs.\ref{fig_arka_SI_3}(h)} and \textbf{(i)}, we plot the total Seebeck coefficient ($S_{xx}$) and the total Nernst coefficient ($S_{yx}$), respectively, at various temperatures (indicated by different colors). First, as the majority carrier type switches at the charge neutrality point (CNP), $S_{xx}$ undergoes a sharp sign change. The steepness of this zero-crossing visibly decreases as the temperature increases. Near the vHS ($n_{0}=\pm 0.75$) at low temperatures, $S_{xx}$ exhibits a small peak that almost entirely vanishes at higher temperatures. Additionally, the Seebeck signal exhibits further temperature dependent sign changes. Regarding the Nernst signal, we observe a prominent, sharp peak located exactly at the van Hove singularity (vHS), which undergoes the expected thermal smearing as the temperature increases. To understand the physical origin of these features near the vHS, we return to the standard definitions of the Seebeck and Nernst coefficients, incorporating the simplifications provided by the Mott relations.
\begin{eqnarray}
\displaystyle
\displaystyle S_{xx}=\frac{\sigma_{xx}\alpha_{xx}+\sigma_{xy}\alpha_{xy}}{\sigma_{xx}^{2}+\sigma_{xy}^{2}}\sim \frac{T}{\Delta}\left(\frac{\partial \sigma_{xx}^{2}}{\partial \varepsilon}+\frac{\partial \sigma_{xy}^{2}}{\partial \varepsilon}\right)_{\varepsilon_{F}}\sim \frac{T}{\Delta}\left(\frac{\partial \Delta}{\partial n}\right) D(\mathcal{E}_{F})
\\
\displaystyle S_{yx}= \frac{\sigma_{xx}\alpha_{yx}-\sigma_{yx}\alpha_{xx}}{\sigma_{xx}^{2}+\sigma_{xy}^{2}}\sim\frac{T}{\Delta}\left(-\sigma_{xx}\frac{\partial \sigma_{xy}}{\partial n}+\sigma_{xy}\frac{\partial\sigma_{xx}}{\partial n}\right) D(\mathcal{E}_{F})
\label{see-arka}
\end{eqnarray}
This expression explicitly demonstrates that the Nernst signal is directly proportional to the density of states. Consequently, it produces a pronounced peak exactly at the vHS, perfectly mirroring the logarithmic divergence of the DOS. For the $S_{xx}$ expression, the prefactor contains the carrier density derivative of the total conductivity magnitude, $\Delta=\sigma_{xx}^{2}+\sigma_{xy}^{2}$. At any point where this magnitude reaches a local extremum (undergoing a change in slope), the Seebeck signal changes sign, passing through a zero-crossing. Note that the model used here has the vHSs but does not represent the tDBLG band structure; therefore, one should not compare the magnitude with the experimental results, but rather the trend in the calculation.
\par

\subsection{Origin of Asymmetry in Nernst Peaks\label{MIN_MODEL_ARKAa}}
\begin{figure*}[htbp]
\centerline{\includegraphics[width=1\textwidth]{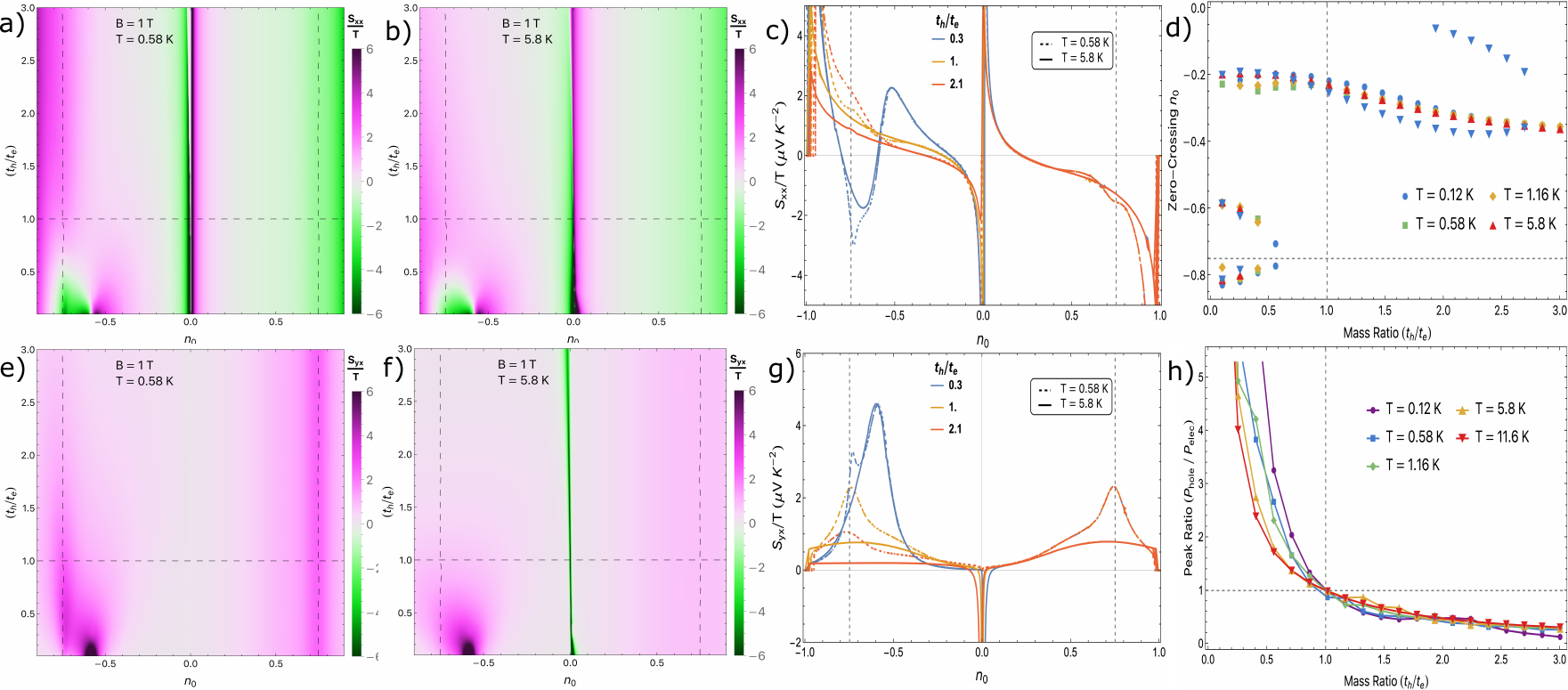}}
\caption{\textbf{Effect of the mass ratio ($t_{h}/t_{e}$) on Seebeck and Nernst coefficients.} \textbf{a}, \textbf{b}, Color maps of the Seebeck coefficient as a function of filling fraction ($n_0$) and mass ratio at low (\textbf{a}) and high (\textbf{b}) temperatures. The color scale indicates positive (pink) and negative (green) values, with zero-crossing points in white. \textbf{c}, Line cuts of the Seebeck coefficient for selected mass ratios ($t_{h}/t_{e} = 0.3, 1, 2$). Solid and dotted lines correspond to high and low temperatures, respectively. \textbf{d}, Trajectories of the Seebeck zero-crossing points across different temperatures. \textbf{e}, \textbf{f}, Color maps of the Nernst coefficient for varying mass ratios and filling fractions. Dark pink regions indicate peaks in the Nernst signal; the green region in \textbf{f} highlights parameter regimes where the high-temperature Nernst coefficient becomes negative. \textbf{g}, Line cuts of the Nernst coefficient for mass ratios 0.3, 1.0, and 2.0 at various temperatures. \textbf{h}, Peak height ratio as a function of the mass ratio, demonstrating a collapse of the curves across different temperatures.}
\label{Sifig5}
\end{figure*}

\noindent Next, we investigate the effect of a varying effective mass ratio ($t_{h}/t_{e}$) on the transport coefficients. Experimental observations indicate an asymmetry in the heights of the Nernst peaks; here, we demonstrate how this variation arises naturally from a mass imbalance between the bands. The first row of SI-Fig.~\ref{Sifig5} presents the Seebeck coefficient in a color plot as a function of the filling fraction ($n_0$) and mass ratio at low ($T=0.58$ K) and high ($T=5.8$ K) temperatures. At lower mass ratios, we observe distinct positive (pink) and negative (green) regions for hole doping ($n_{0} < 0$). However, as the mass ratio increases, this regime becomes entirely positive regardless of temperature. SI-Fig.~\ref{Sifig5} (c) highlights specific line cuts at $t_h/t_e = 0.3$, $1.0$, and $2.1$. Notably, for smaller mass ratios, the Seebeck coefficient exhibits multiple zero-crossings, whereas for larger mass ratios, no zero-crossings occur in the vicinity of the van Hove singularity (vHS). Finally, SI-Fig.~\ref{Sifig5}d tracks the evolution of these zero-crossing points across different temperatures and mass ratios, revealing the emergence of a robust zero-crossing near $n_{0}=0$. here near $n_{0}=-0.75$ we can see a converging pattern along with another transition curve starting at $n_{0}=-0.2$.

\noindent The second row illustrates the Nernst coefficient as a function of mass ratio and filling fraction at low (e) and high (f) temperatures. For smaller mass ratios, a distinct primary peak emerges near the van Hove singularity (vHS), accompanied by a secondary peak. As the mass ratio increases, this secondary feature gradually vanishes. Throughout this regime, the low-temperature Nernst signal remains strictly positive. Conversely, at higher temperatures (f), the Nernst coefficient becomes negative near the $n_{0}=0$ point across all mass ratios. Line cuts for selected mass ratios ($t_h/t_e = 0.3, 1.0, 2.1$) at various temperatures are shown in (g). A distinct Nernst peak remains visible near the vHS in all cases, although significant smearing occurs at larger mass ratios. Finally, in (h), we plot the ratio of the maximum Nernst values between the electron-doped ($n_{0}>0$) and hole-doped ($n_{0}<0$) regimes. Remarkably, the curves for different temperatures collapse onto a single universal curve, demonstrating a clear decreasing trend as the mass ratio increases.


\noindent\textbf{References}
\putbib[ref_si]

\end{bibunit}

\begin{thebibliography}{10}
\expandafter\ifx\csname url\endcsname\relax
  \def\url#1{\texttt{#1}}\fi
\expandafter\ifx\csname urlprefix\endcsname\relax\def\urlprefix{URL }\fi
\providecommand{\bibinfo}[2]{#2}
\providecommand{\eprint}[2][]{\url{#2}}

\bibitem{Andrei2021}
\bibinfo{author}{Andrei, E.~Y.} \emph{et~al.}
\newblock \bibinfo{title}{The marvels of moir{\'e} materials}.
\newblock \emph{\bibinfo{journal}{Nat. Rev. Mater.}} \textbf{\bibinfo{volume}{6}}, \bibinfo{pages}{201--206} (\bibinfo{year}{2021}).

\bibitem{Andrei2020-ad}
\bibinfo{author}{Andrei, E.~Y.} \& \bibinfo{author}{MacDonald, A.~H.}
\newblock \bibinfo{title}{Graphene bilayers with a twist}.
\newblock \emph{\bibinfo{journal}{Nat. Mater.}} \textbf{\bibinfo{volume}{19}}, \bibinfo{pages}{1265--1275} (\bibinfo{year}{2020}).

\bibitem{Jung2019}
\bibinfo{author}{Chebrolu, N.~R.}, \bibinfo{author}{Chittari, B.~L.} \& \bibinfo{author}{Jung, J.}
\newblock \bibinfo{title}{Flat bands in twisted double bilayer graphene}.
\newblock \emph{\bibinfo{journal}{Phys. Rev. B.}} \textbf{\bibinfo{volume}{99}} (\bibinfo{year}{2019}).

\bibitem{Koshino2019}
\bibinfo{author}{Koshino, M.}
\newblock \bibinfo{title}{Band structure and topological properties of twisted double bilayer graphene}.
\newblock \emph{\bibinfo{journal}{Phys. Rev. B.}} \textbf{\bibinfo{volume}{99}} (\bibinfo{year}{2019}).

\bibitem{Cao2020}
\bibinfo{author}{Cao, Y.} \emph{et~al.}
\newblock \bibinfo{title}{Tunable correlated states and spin-polarized phases in twisted bilayer-bilayer graphene}.
\newblock \emph{\bibinfo{journal}{Nature}} \textbf{\bibinfo{volume}{583}}, \bibinfo{pages}{215--220} (\bibinfo{year}{2020}).

\bibitem{PKim2020}
\bibinfo{author}{Liu, X.} \emph{et~al.}
\newblock \bibinfo{title}{Tunable spin-polarized correlated states in twisted double bilayer graphene}.
\newblock \emph{\bibinfo{journal}{Nature}} \textbf{\bibinfo{volume}{583}}, \bibinfo{pages}{221--225} (\bibinfo{year}{2020}).

\bibitem{Shen2020}
\bibinfo{author}{Shen, C.} \emph{et~al.}
\newblock \bibinfo{title}{Correlated states in twisted double bilayer graphene}.
\newblock \emph{\bibinfo{journal}{Nat. Phys.}} \textbf{\bibinfo{volume}{16}}, \bibinfo{pages}{520--525} (\bibinfo{year}{2020}).

\bibitem{yankowitz2021}
\bibinfo{author}{He, M.} \emph{et~al.}
\newblock \bibinfo{title}{Symmetry breaking in twisted double bilayer graphene}.
\newblock \emph{\bibinfo{journal}{Nat. Phys.}} \textbf{\bibinfo{volume}{17}}, \bibinfo{pages}{26--30} (\bibinfo{year}{2021}).

\bibitem{Su2023}
\bibinfo{author}{Su, R.}, \bibinfo{author}{Kuiri, M.}, \bibinfo{author}{Watanabe, K.}, \bibinfo{author}{Taniguchi, T.} \& \bibinfo{author}{Folk, J.}
\newblock \bibinfo{title}{Superconductivity in twisted double bilayer graphene stabilized by {WSe2}}.
\newblock \emph{\bibinfo{journal}{Nat. Mater.}} \textbf{\bibinfo{volume}{22}}, \bibinfo{pages}{1332--1337} (\bibinfo{year}{2023}).

\bibitem{Ashvin2019}
\bibinfo{author}{Lee, J.~Y.} \emph{et~al.}
\newblock \bibinfo{title}{Theory of correlated insulating behaviour and spin-triplet superconductivity in twisted double bilayer graphene}.
\newblock \emph{\bibinfo{journal}{Nat. Commun.}} \textbf{\bibinfo{volume}{10}}, \bibinfo{pages}{5333} (\bibinfo{year}{2019}).

\bibitem{Yazdani2020}
\bibinfo{author}{Wong, D.} \emph{et~al.}
\newblock \bibinfo{title}{Cascade of electronic transitions in magic-angle twisted bilayer graphene}.
\newblock \emph{\bibinfo{journal}{Nature}} \textbf{\bibinfo{volume}{582}}, \bibinfo{pages}{198--202} (\bibinfo{year}{2020}).

\bibitem{Zondiner2020}
\bibinfo{author}{Zondiner, U.} \emph{et~al.}
\newblock \bibinfo{title}{Cascade of phase transitions and dirac revivals in magic-angle graphene}.
\newblock \emph{\bibinfo{journal}{Nature}} \textbf{\bibinfo{volume}{582}}, \bibinfo{pages}{203--208} (\bibinfo{year}{2020}).

\bibitem{Kohn1965}
\bibinfo{author}{Kohn, W.} \& \bibinfo{author}{Luttinger, J.~M.}
\newblock \bibinfo{title}{New mechanism for superconductivity}.
\newblock \emph{\bibinfo{journal}{Phys. Rev. Lett.}} \textbf{\bibinfo{volume}{15}}, \bibinfo{pages}{524--526} (\bibinfo{year}{1965}).

\bibitem{Gonzalez2008-rw}
\bibinfo{author}{Gonz{\'a}lez, J.}
\newblock \bibinfo{title}{{Kohn-Luttinger} superconductivity in graphene}.
\newblock \emph{\bibinfo{journal}{Phys. Rev. B Condens. Matter Mater. Phys.}} \textbf{\bibinfo{volume}{78}} (\bibinfo{year}{2008}).

\bibitem{Nandkishore2012-cv}
\bibinfo{author}{Nandkishore, R.}, \bibinfo{author}{Levitov, L.~S.} \& \bibinfo{author}{Chubukov, A.~V.}
\newblock \bibinfo{title}{Chiral superconductivity from repulsive interactions in doped graphene}.
\newblock \emph{\bibinfo{journal}{Nat. Phys.}} \textbf{\bibinfo{volume}{8}}, \bibinfo{pages}{158--163} (\bibinfo{year}{2012}).

\bibitem{Fleck1997-yo}
\bibinfo{author}{Fleck, M.}, \bibinfo{author}{Ole{\'s}, A.~M.} \& \bibinfo{author}{Hedin, L.}
\newblock \bibinfo{title}{Magnetic phases near the van hove singularity ins- andd-band hubbard models}.
\newblock \emph{\bibinfo{journal}{Phys. Rev. B Condens. Matter}} \textbf{\bibinfo{volume}{56}}, \bibinfo{pages}{3159--3166} (\bibinfo{year}{1997}).

\bibitem{Betouras2018}
\bibinfo{author}{Sherkunov, Y.} \& \bibinfo{author}{Betouras, J.~J.}
\newblock \bibinfo{title}{Electronic phases in twisted bilayer graphene at magic angles as a result of van hove singularities and interactions}.
\newblock \emph{\bibinfo{journal}{Phys. Rev. B}} \textbf{\bibinfo{volume}{98}}, \bibinfo{pages}{205151} (\bibinfo{year}{2018}).
\newblock \urlprefix\url{https://link.aps.org/doi/10.1103/PhysRevB.98.205151}.

\bibitem{He2019}
\bibinfo{author}{Liu, Y.-W.} \emph{et~al.}
\newblock \bibinfo{title}{Magnetism near half-filling of a van hove singularity in twisted graphene bilayer}.
\newblock \emph{\bibinfo{journal}{Phys. Rev. B}} \textbf{\bibinfo{volume}{99}}, \bibinfo{pages}{201408(R)} (\bibinfo{year}{2019}).
\newblock \urlprefix\url{https://link.aps.org/doi/10.1103/PhysRevB.99.201408}.

\bibitem{Andrei2010}
\bibinfo{author}{Li, G.} \emph{et~al.}
\newblock \bibinfo{title}{Observation of van hove singularities in twisted graphene layers}.
\newblock \emph{\bibinfo{journal}{Nat. Phys.}} \textbf{\bibinfo{volume}{6}}, \bibinfo{pages}{109--113} (\bibinfo{year}{2010}).

\bibitem{Pasupathy2019}
\bibinfo{author}{Kerelsky, A.} \emph{et~al.}
\newblock \bibinfo{title}{Maximized electron interactions at the magic angle in twisted bilayer graphene}.
\newblock \emph{\bibinfo{journal}{Nature}} \textbf{\bibinfo{volume}{572}}, \bibinfo{pages}{95--100} (\bibinfo{year}{2019}).

\bibitem{Nadj2019}
\bibinfo{author}{Choi, Y.} \emph{et~al.}
\newblock \bibinfo{title}{Electronic correlations in twisted bilayer graphene near the magic angle}.
\newblock \emph{\bibinfo{journal}{Nat. Phys.}} \textbf{\bibinfo{volume}{15}}, \bibinfo{pages}{1174--1180} (\bibinfo{year}{2019}).

\bibitem{Wu2021}
\bibinfo{author}{Wu, S.}, \bibinfo{author}{Zhang, Z.}, \bibinfo{author}{Watanabe, K.}, \bibinfo{author}{Taniguchi, T.} \& \bibinfo{author}{Andrei, E.~Y.}
\newblock \bibinfo{title}{Chern insulators, van hove singularities and topological flat bands in magic-angle twisted bilayer graphene}.
\newblock \emph{\bibinfo{journal}{Nat. Mater.}} \textbf{\bibinfo{volume}{20}}, \bibinfo{pages}{488--494} (\bibinfo{year}{2021}).

\bibitem{Mak2025}
\bibinfo{author}{Kn{\"u}ppel, P.} \emph{et~al.}
\newblock \bibinfo{title}{Correlated states controlled by a tunable van hove singularity in moir{\'e} {WSe2} bilayers}.
\newblock \emph{\bibinfo{journal}{Nat. Commun.}} \textbf{\bibinfo{volume}{16}}, \bibinfo{pages}{1959} (\bibinfo{year}{2025}).

\bibitem{Behnia2009}
\bibinfo{author}{Behnia, K.}
\newblock \bibinfo{title}{The nernst effect and the boundaries of the fermi liquid picture}.
\newblock \emph{\bibinfo{journal}{J. Phys. Condens. Matter}} \textbf{\bibinfo{volume}{21}}, \bibinfo{pages}{113101} (\bibinfo{year}{2009}).

\bibitem{Behnia2016-es}
\bibinfo{author}{Behnia, K.} \& \bibinfo{author}{Aubin, H.}
\newblock \bibinfo{title}{Nernst effect in metals and superconductors: a review of concepts and experiments}.
\newblock \emph{\bibinfo{journal}{Rep. Prog. Phys.}} \textbf{\bibinfo{volume}{79}}, \bibinfo{pages}{046502} (\bibinfo{year}{2016}).

\bibitem{Sondheimer1948-ff}
\bibinfo{author}{Sondheimer, E.~H.}
\newblock \bibinfo{title}{The theory of the galvanomagnetic and thermomagnetic effects in metals}.
\newblock \emph{\bibinfo{journal}{Proc. R. Soc. Lond.}} \textbf{\bibinfo{volume}{193}}, \bibinfo{pages}{484--512} (\bibinfo{year}{1948}).
\newblock \urlprefix\url{https://royalsocietypublishing.org/rspa/article/193/1035/484/6698/The-theory-of-the-galvanomagnetic-and}.

\bibitem{Behnia2007}
\bibinfo{author}{Behnia, K.}, \bibinfo{author}{M\'easson, M.-A.} \& \bibinfo{author}{Kopelevich, Y.}
\newblock \bibinfo{title}{Nernst effect in semimetals: The effective mass and the figure of merit}.
\newblock \emph{\bibinfo{journal}{Phys. Rev. Lett.}} \textbf{\bibinfo{volume}{98}}, \bibinfo{pages}{076603} (\bibinfo{year}{2007}).
\newblock \urlprefix\url{https://link.aps.org/doi/10.1103/PhysRevLett.98.076603}.

\bibitem{Ong2000-os}
\bibinfo{author}{Xu, Z.~A.}, \bibinfo{author}{Ong, N.~P.}, \bibinfo{author}{Wang, Y.}, \bibinfo{author}{Kakeshita, T.} \& \bibinfo{author}{Uchida, S.}
\newblock \bibinfo{title}{Vortex-like excitations and the onset of superconducting phase fluctuation in underdoped {La(2-x)Sr(x)CuO4}}.
\newblock \emph{\bibinfo{journal}{Nature}} \textbf{\bibinfo{volume}{406}}, \bibinfo{pages}{486--488} (\bibinfo{year}{2000}).
\newblock \urlprefix\url{https://www.nature.com/articles/35020016}.

\bibitem{ong2006}
\bibinfo{author}{Wang, Y.}, \bibinfo{author}{Li, L.} \& \bibinfo{author}{Ong, N.~P.}
\newblock \bibinfo{title}{Nernst effect in high-${T}_{c}$ superconductors}.
\newblock \emph{\bibinfo{journal}{Phys. Rev. B}} \textbf{\bibinfo{volume}{73}}, \bibinfo{pages}{024510} (\bibinfo{year}{2006}).
\newblock \urlprefix\url{https://link.aps.org/doi/10.1103/PhysRevB.73.024510}.

\bibitem{Pourret2006-pq}
\bibinfo{author}{Pourret, A.} \emph{et~al.}
\newblock \bibinfo{title}{Observation of the nernst signal generated by fluctuating cooper pairs}.
\newblock \emph{\bibinfo{journal}{Nat. Phys.}} \textbf{\bibinfo{volume}{2}}, \bibinfo{pages}{683--686} (\bibinfo{year}{2006}).
\newblock \urlprefix\url{https://www.nature.com/articles/nphys413}.

\bibitem{Huse2002}
\bibinfo{author}{Ussishkin, I.}, \bibinfo{author}{Sondhi, S.~L.} \& \bibinfo{author}{Huse, D.~A.}
\newblock \bibinfo{title}{Gaussian superconducting fluctuations, thermal transport, and the nernst effect}.
\newblock \emph{\bibinfo{journal}{Phys. Rev. Lett.}} \textbf{\bibinfo{volume}{89}}, \bibinfo{pages}{287001} (\bibinfo{year}{2002}).
\newblock \urlprefix\url{https://link.aps.org/doi/10.1103/PhysRevLett.89.287001}.

\bibitem{Behnia2010}
\bibinfo{author}{Zhu, Z.}, \bibinfo{author}{Yang, H.}, \bibinfo{author}{Fauqu{\'e}, B.}, \bibinfo{author}{Kopelevich, Y.} \& \bibinfo{author}{Behnia, K.}
\newblock \bibinfo{title}{Nernst effect and dimensionality in the quantum limit}.
\newblock \emph{\bibinfo{journal}{Nat. Phys.}} \textbf{\bibinfo{volume}{6}}, \bibinfo{pages}{26--29} (\bibinfo{year}{2010}).
\newblock \urlprefix\url{https://www.nature.com/articles/nphys1437}.

\bibitem{Behnia2015wte2}
\bibinfo{author}{Zhu, Z.} \emph{et~al.}
\newblock \bibinfo{title}{Quantum oscillations, thermoelectric coefficients, and the fermi surface of semimetallic ${\mathrm{wte}}_{2}$}.
\newblock \emph{\bibinfo{journal}{Phys. Rev. Lett.}} \textbf{\bibinfo{volume}{114}}, \bibinfo{pages}{176601} (\bibinfo{year}{2015}).
\newblock \urlprefix\url{https://link.aps.org/doi/10.1103/PhysRevLett.114.176601}.

\bibitem{Ong2013}
\bibinfo{author}{Liang, T.} \emph{et~al.}
\newblock \bibinfo{title}{Evidence for massive bulk dirac fermions in pb1−xsnxse from nernst and thermopower experiments}.
\newblock \emph{\bibinfo{journal}{Nat. Commun.}} \textbf{\bibinfo{volume}{4}}, \bibinfo{pages}{2696} (\bibinfo{year}{2013}).
\newblock \urlprefix\url{https://www.nature.com/articles/ncomms3696}.

\bibitem{Behnia2015}
\bibinfo{author}{Fauqu\'e, B.} \emph{et~al.}
\newblock \bibinfo{title}{Magnetothermoelectric properties of bi${}_{2}$se${}_{3}$}.
\newblock \emph{\bibinfo{journal}{Phys. Rev. B}} \textbf{\bibinfo{volume}{87}}, \bibinfo{pages}{035133} (\bibinfo{year}{2013}).
\newblock \urlprefix\url{https://link.aps.org/doi/10.1103/PhysRevB.87.035133}.

\bibitem{BiTei2015}
\bibinfo{author}{Ideue, T.} \emph{et~al.}
\newblock \bibinfo{title}{Thermoelectric probe for fermi surface topology in the three-dimensional rashba semiconductor bitei}.
\newblock \emph{\bibinfo{journal}{Phys. Rev. B}} \textbf{\bibinfo{volume}{92}}, \bibinfo{pages}{115144} (\bibinfo{year}{2015}).
\newblock \urlprefix\url{https://link.aps.org/doi/10.1103/PhysRevB.92.115144}.

\bibitem{Behnia2003}
\bibinfo{author}{Bel, R.}, \bibinfo{author}{Behnia, K.} \& \bibinfo{author}{Berger, H.}
\newblock \bibinfo{title}{Ambipolar nernst effect in ${\mathrm{n}\mathrm{b}\mathrm{s}\mathrm{e}}_{2}$}.
\newblock \emph{\bibinfo{journal}{Phys. Rev. Lett.}} \textbf{\bibinfo{volume}{91}}, \bibinfo{pages}{066602} (\bibinfo{year}{2003}).
\newblock \urlprefix\url{https://link.aps.org/doi/10.1103/PhysRevLett.91.066602}.

\bibitem{sanfengWu2016}
\bibinfo{author}{Wu, S.} \emph{et~al.}
\newblock \bibinfo{title}{Multiple hot-carrier collection in photo-excited graphene moir{\'e} superlattices}.
\newblock \emph{\bibinfo{journal}{Sci. Adv.}} \textbf{\bibinfo{volume}{2}}, \bibinfo{pages}{e1600002} (\bibinfo{year}{2016}).
\newblock \urlprefix\url{https://www.science.org/doi/10.1126/sciadv.1600002}.

\bibitem{Mohan2021}
\bibinfo{author}{Mohan, P.}, \bibinfo{author}{Ghorai, U.} \& \bibinfo{author}{Sensarma, R.}
\newblock \bibinfo{title}{Trigonal warping, satellite dirac points, and multiple field tuned topological transitions in twisted double bilayer graphene}.
\newblock \emph{\bibinfo{journal}{Phys. Rev. B}} \textbf{\bibinfo{volume}{103}}, \bibinfo{pages}{155149} (\bibinfo{year}{2021}).
\newblock \urlprefix\url{https://link.aps.org/doi/10.1103/PhysRevB.103.155149}.

\bibitem{kim2009}
\bibinfo{author}{Zuev, Y.~M.}, \bibinfo{author}{Chang, W.} \& \bibinfo{author}{Kim, P.}
\newblock \bibinfo{title}{Thermoelectric and magnetothermoelectric transport measurements of graphene}.
\newblock \emph{\bibinfo{journal}{Phys. Rev. Lett.}} \textbf{\bibinfo{volume}{102}}, \bibinfo{pages}{096807} (\bibinfo{year}{2009}).
\newblock \urlprefix\url{https://link.aps.org/doi/10.1103/PhysRevLett.102.096807}.

\bibitem{Paul2022-sn}
\bibinfo{author}{Paul, A.~K.} \emph{et~al.}
\newblock \bibinfo{title}{Interaction-driven giant thermopower in magic-angle twisted bilayer graphene}.
\newblock \emph{\bibinfo{journal}{Nat. Phys.}} \textbf{\bibinfo{volume}{18}}, \bibinfo{pages}{691--698} (\bibinfo{year}{2022}).
\newblock \urlprefix\url{https://www.nature.com/articles/s41567-022-01574-3}.

\bibitem{Mott2012-kk}
\bibinfo{author}{Mott, N.~F.} \& \bibinfo{author}{Davis, E.~A.}
\newblock \emph{\bibinfo{title}{Electronic processes in non-crystalline materials}}.
\newblock Oxford Classic Texts in the Physical Sciences (\bibinfo{publisher}{Oxford University Press}, \bibinfo{address}{London, England}, \bibinfo{year}{2012}).

\bibitem{Ussishkin2004}
\bibinfo{author}{Oganesyan, V.} \& \bibinfo{author}{Ussishkin, I.}
\newblock \bibinfo{title}{Nernst effect, quasiparticles, and $d$-density waves in cuprates}.
\newblock \emph{\bibinfo{journal}{Phys. Rev. B}} \textbf{\bibinfo{volume}{70}}, \bibinfo{pages}{054503} (\bibinfo{year}{2004}).
\newblock \urlprefix\url{https://link.aps.org/doi/10.1103/PhysRevB.70.054503}.

\end{thebibliography}


\begin{thebibliography}{1}
\expandafter\ifx\csname url\endcsname\relax
  \def\url#1{\texttt{#1}}\fi
\expandafter\ifx\csname urlprefix\endcsname\relax\def\urlprefix{URL }\fi
\providecommand{\bibinfo}[2]{#2}
\providecommand{\eprint}[2][]{\url{#2}}

\bibitem{Koshino2019}
\bibinfo{author}{Koshino, M.}
\newblock \bibinfo{title}{Band structure and topological properties of twisted double bilayer graphene}.
\newblock \emph{\bibinfo{journal}{Phys. Rev. B.}} \textbf{\bibinfo{volume}{99}} (\bibinfo{year}{2019}).

\bibitem{Mohan2021}
\bibinfo{author}{Mohan, P.}, \bibinfo{author}{Ghorai, U.} \& \bibinfo{author}{Sensarma, R.}
\newblock \bibinfo{title}{Trigonal warping, satellite dirac points, and multiple field tuned topological transitions in twisted double bilayer graphene}.
\newblock \emph{\bibinfo{journal}{Phys. Rev. B}} \textbf{\bibinfo{volume}{103}}, \bibinfo{pages}{155149} (\bibinfo{year}{2021}).
\newblock \urlprefix\url{https://link.aps.org/doi/10.1103/PhysRevB.103.155149}.

\bibitem{Behnia2009}
\bibinfo{author}{Behnia, K.}
\newblock \bibinfo{title}{The nernst effect and the boundaries of the fermi liquid picture}.
\newblock \emph{\bibinfo{journal}{J. Phys. Condens. Matter}} \textbf{\bibinfo{volume}{21}}, \bibinfo{pages}{113101} (\bibinfo{year}{2009}).

\end{thebibliography}
\end{document}